\documentclass{aa}

\usepackage{graphicx}
\usepackage{txfonts}
\usepackage{siunitx}
\usepackage{hyperref}
\bibpunct{(}{)}{;}{a}{}{,} 

\begin{document}

   \title{A global model of the magnetorotational instability in protoneutron stars}

   \subtitle{}

   \author{A. Reboul-Salze\thanks{\email{alexis.reboul-salze@cea.fr}
}
          \inst{}
          \and
          J. Guilet\inst{}
          \and
          R. Raynaud\inst{}
          \and
          M. Bugli\inst{}
   }

   \institute{Laboratoire AIM, CEA/DRF-CNRS-Universit\'e Paris Diderot, IRFU/
              D\'epartement d'Astrophysique, CEA-Saclay, F-91191, France\\
              \email{alexis.reboul-salze@cea.fr}
             }

   \date{Received 07 May 2020; Accepted 18 October 2020}


  \abstract
  {Magnetars are isolated neutron stars characterized by their variable high-energy emission, which is powered by the dissipation of enormous internal magnetic fields. The measured spin-down of magnetars constrains the magnetic dipole to be in the range of $10^{14}$ to $10^{15}$ G. The magnetorotational instability (MRI) is considered to be a promising mechanism to amplify the magnetic field in fast-rotating protoneutron stars and form magnetars. This scenario is supported by many local studies that have shown that magnetic fields could be amplified by the MRI on small scales. However, the efficiency of the MRI at generating a dipole field is still unknown.}
  {To answer this question, we study the MRI dynamo in an idealized global model of a fast rotating protoneutron star with differential rotation.}
  {Using the pseudo-spectral code MagIC, we performed three-dimensional incompressible magnetohydrodynamics simulations in spherical geometry with explicit diffusivities where the differential rotation is forced at the outer boundary. We performed a parameter study in which we varied the initial magnetic field and investigated different magnetic boundary conditions.
   These simulations were compared to local shearing box simulations performed with the code Snoopy.}
  {We obtain a self-sustained turbulent MRI-driven dynamo, whose saturated state is independent of the initial magnetic field. The MRI generates a strong turbulent magnetic field of $B \geq 2\times 10^{15} \ \mathrm{G}$ and a nondominant magnetic dipole, which represents systematically about $5\%$ of the averaged magnetic field strength.
  Interestingly, this dipole is tilted toward the equatorial plane.
  By comparing these results with shearing box simulations, we find that local models can reproduce fairly well several characteristics of global MRI turbulence such as the kinetic and magnetic spectra. The turbulence is nonetheless more vigorous in the local models than in the global ones. Moreover, overly large boxes allow for elongated structures to develop without any realistic curvature constraint, which may explain why these models tend to overestimate the field amplification.}
  {Overall, our results support the ability of the MRI to form magnetar-like large-scale magnetic fields. They furthermore predict the presence of a stronger small-scale magnetic field. The resulting magnetic field could be important to power outstanding stellar explosions, such as superluminous supernovae and gamma-ray bursts.}

\maketitle
%

\section{Introduction}

Magnetars are a class of young and highly magnetized neutron stars that produce a wide variety of outstanding emission at X-ray or soft $\gamma$-ray energies \citep[and references therein]{1998Natur.393..235K,2017ARA&A..55..261K}.
The period and the spin-down of these objects are measured by the long time follow-up of their pulsed X-ray activity, and a surface dipolar magnetic field of $B_\mathrm{dip} =  10^{14}\left( \frac{P}{5\ \mathrm{s}}\right) ^{\frac{1}{2}} \left( \frac{\dot{P}}{10^{-11}\ \mathrm{s} \ \mathrm{s}^{-1}}\right) ^{\frac{1}{2}} \mathrm{G} \simeq 10^{14}$--$10^{15} \ \mathrm{G}$ can be inferred under the assumption of magnetic dipole spin-down \citep{2014ApJS..212....6O}\footnote{\url{http://www.physics.mcgill.ca/~pulsar/magnetar/main.html}}.
Their activity can be explained by the decay of their ultrastrong magnetic field and also includes short bursts \citep{2006A&A...445..313G}, large outbursts \citep{2018MNRAS.474..961C}, giant flares \citep{2005Natur.434.1098H}, and quasi-periodic oscillations \citep{2005ApJ...628L..53I}.
Absorption lines are also been detected in outbursts for two objects and interpreted as proton cyclotron lines \citep{2013Natur.500..312T,2016MNRAS.456.4145R}. This suggests the presence of a strong non-dipolar surface field, whereas a weaker dipolar component is derived from the timing parameters of these objects.

Magnetic fields, especially in the presence of fast rotation, could play an important role in the dynamics of core-collapse supernovae and have garnered considerable interest in the last decade.
Indeed, a strong magnetic field would efficiently extract the large rotational energy of a protoneutron star (PNS) rotating with a period of a few milliseconds.
The presence of a magnetic field can impact the explosion by converting the energy of differential rotation into thermal energy \citep{2005ApJ...620..861T} and/or into a large-scale magnetic field, which can launch jets and lead to a magnetorotational explosion \citep{2006MNRAS.370..501M,2006PhRvD..74j4026S,2008ApJ...673L..43D,2012ApJ...750L..22W,2014ApJ...785L..29M,2018JPhG...45h4001O,2020MNRAS.492...58B,2020arXiv200302004K}.
Moreover, a millisecond proto-magnetar is a potential central engine for long gamma-ray bursts \citep{1992ApJ...392L...9D,2011MNRAS.413.2031M,2018ApJ...857...95M} associated with ``hypernovae'' or supernovae type Ic Broad Lined \citep{2011ApJ...741...97D}. These rare events are characterized by a kinetic energy ten times higher than that of standard supernovae. Furthermore, millisecond magnetars have been invoked to explain some superluminous supernovae through a delayed energy injection due to the dipole spin-down luminosity \citep{2013Natur.502..346N,2013ApJ...770..128I,2018MNRAS.475.2659M}. Millisecond magnetars may also be formed in binary neutron star mergers. This can provide a natural explanation for the extended emission and the plateau phase in X-ray sources associated with a fraction of short gamma-ray bursts \citep{2008MNRAS.385.1455M,2012MNRAS.419.1537B,2013MNRAS.430.1061R,2014MNRAS.438..240G}. An X-ray transient has also recently been detected and interpreted as the formation of a magnetar in the aftermath of a binary neutron-star merger \citep{2019Natur.568..198X}.

Several scenarios have been invoked to explain the origin of the magnetic field in magnetars. It may stem from the field of the progenitor amplified by magnetic flux conservation.
This scenario can lead to the strongest magnetic fields in the case of highly magnetized progenitors that could be formed in stellar mergers \citep{2019Natur.574..211S}, although it may not explain the formation of millisecond magnetars since highly magnetized progenitors are slow rotators \citep{2008AIPC..983..391S,2018MNRAS.475.5144S}.
An alternative process is an in situ magnetic field amplification by a turbulent dynamo in the PNS, either by a convective dynamo \citep{1993ApJ...408..194T,2020ScienceRaynaud} or the magnetorotational instability \citep[MRI, see][]{2003ApJ...584..954A,2009A&A...498..241O}.

The first local analytical study of the MRI in the context of Keplerian accretion disks by \citet{1991ApJ...376..214B} showed that in presence of differential rotation small seed perturbations are amplified exponentially with time. The first local simulations by \citet{1992ApJ...400..595H} showed that the turbulent velocity and magnetic field reach a statistically stationary state.
In ideal magnetohydrodynamics (MHD) and with an initial vertical uniform magnetic field, the growth rate of the instability is on the order of the rotation rate $\Omega$. In this case, the wavelength of the fastest growing mode is proportional to the magnetic field intensity.
Therefore, the weaker the magnetic field, the shorter the MRI wavelength and the more difficult it becomes to resolve in a global model.
This instability has thus widely been studied in the local approximation, either analytically or by using ``shearing box'' simulations representing a part of the accretion disk.
The linear growth of the MRI has been studied with thermal stratification (entropy and composition gradient) and with diffusion processes (viscosity and resistivity) \citep{1995ApJ...453..380B,2004ApJ...607..564M,2008ApJ...684..498P}.

Core-collapse supernova simulations show a strong differential rotation in the PNS \citep{2003ApJ...584..954A,2006ApJS..164..130O}.
Numerical models in the context of supernovae have shown that the MRI can grow on shorter timescales than the successful explosion time and that an efficient amplification of the magnetic field occurs at small scales \citep{2009A&A...498..241O,2015MNRAS.450.2153G,2016MNRAS.460.3316R}.
The influence of the specific physical conditions of PNSs were studied in these local models.
First, the pressure gradient rather than the centrifugal force balances gravity, which can lead to a non-Keplerian rotation profile.
In the case of a steeper rotation profile, a stronger MRI turbulence develops in shearing box simulations \citep{2012ApJ...759..110M}.
Secondly, due to the high density inside the PNS, neutrinos are in the diffusive regime and their transport of momentum can be described with a high viscosity, which can limit the growth of the MRI if the initial magnetic field is too low \citep{2015MNRAS.447.3992G}.
Finally, the buoyancy forces driven by the entropy and lepton fraction can reduce the MRI turbulence in the case of stable stratification \citep{2015MNRAS.450.2153G}.

The impact of the spherical geometry of the full PNS on the MRI turbulence and the ability of the MRI to generate a large-scale field, similar to the inferred magnetic field of magnetars, is still unknown.
The first attempts to address this question rely on semi-global models that include radial gradients of density and entropy \citep{2009A&A...498..241O,2015ApJ...798L..22M}. However, these models remain local at least vertically and therefore cannot investigate the generation of a large-scale magnetic field.
Global axisymmetric simulations of the MRI also show magnetic field amplification \citep{2013ApJ...770L..19S,2016ApJ...817..153S}.
\citet{2015Natur.528..376M} performed the first simulations describing a quarter of the PNS with a high enough resolution to resolve the MRI wavelength and  showed the development of the MRI turbulence.
The model was, however, started with an initial magnetic dipole and therefore did not demonstrate the generation of a magnetar-like magnetic dipole.

This paper studies for the first time the global properties of the MRI in a full 3D spherical model, where no initial large-scale magnetic field is assumed.
To resolve the MRI wavelength with a reasonable resolution, we use a sufficiently strong initial magnetic field.
However, only the small scales are initialized in order to study the generation of the magnetic dipole. This intense and small-scale magnetic field can be interpreted as the result of the first amplification described in local models.
With respect to previous studies, a different approach is also used for the physical setup, which is reduced to its most fundamental ingredients. This has the advantage of providing a useful reference for our physical understanding, while at the same time drastically reducing the computational cost and enabling long simulation times and the exploration of the parameter space.

The paper is organized as follows. In Sect. \ref{Numerical}, we describe the physical and numerical setup. The results are then presented in Sect. \ref{results} for the saturated nonlinear phase of the MRI, and in Sect. \ref{local_comp} for the comparison of our global model to local models of the MRI. Finally, we discuss the validity of our assumptions in Sect. \ref{discuss} and draw our conclusions in Sect. \ref{conclusions}.

\section{Numerical setup}
\label{Numerical}

\subsection{Governing equations}
\label{Eq_intro}
The simulations performed in this article are designed to represent a fast rotating PNS.
We assume that the hot PNS has a mass of $1.3 M_{\odot}$ and a radius of $r_\mathrm{o}=\SI{25}{\km}$.
As the PNS is in solid body rotation for a radius $r \le \SI{10}{\km}$ (see Sect.~\ref{s:ic} for the assumed rotation profile), the inner core of the PNS is stable to the MRI and it can be excluded from the simulation domain. We chose an inner core with a radius $r_\mathrm{i}=\SI{6.25}{\km}$.
The shell gap $D \equiv r_\mathrm{o} - r_\mathrm{i}=\SI{18.75}{\km}$ is the characteristic length of the simulation domain.
In our model, we assume that neutrinos are in the diffusive regime such that their effects on the dynamics
can be appropriately described by a viscosity $\nu$ \citep{2015MNRAS.447.3992G}.
The incompressible approximation is used for the sake of simplicity.
We assume a uniform density $\rho_0=\SI{4e13} {\gram \per \cm \cubed}$ and neglect buoyancy effects.
The diffusive incompressible MHD equations describing the dynamics of the PNS in a rotating frame
at an angular frequency $\Omega_0 = \SI{1000}{\per\second}$ read
\begin{gather}
  \dfrac{\partial \vec{u}}{\partial t}+ \left(\vec{u}\cdot\vec{\nabla}\right)
  \vec{u}  =-\vec{\nabla} p' -2\vec{\Omega_0}\times\vec{u}
  + \dfrac{1}{\mu_0\rho_0}\left(\vec{\nabla}\times\vec{B}\right)\times\vec{B}
  +  \nu \Delta \vec{u} \,, \label{eq:1} \\
  \dfrac{\partial \vec{B}}{\partial t} = \vec{\nabla} \times \left( \vec{u}\times\vec{B}-\eta\,\vec{\nabla}\times\vec{B}\right) \,, \label{eq:2} \\
  \vec{\nabla} \cdot \vec{u} = 0 \,, \label{eq:3} \\
  \vec{\nabla} \cdot \vec{B} = 0 \label{eq:4}\,,
\end{gather}
where $\vec{u}$ is the flow velocity, $\vec{B}$ the magnetic field, $p'$ the reduced pressure (i.e., the pressure normalized by the density) and $\mu_0$ the vacuum permeability.

\subsection{Numerical methods}

In order to solve the incompressible MHD system of equations~\eqref{eq:1}--\eqref{eq:4}, we used the pseudo-spectral code MagIC \citep{2002PEPI..132..281W,GASTINE2012428,2013GGG....14..751S}\footnote{\url{https://magic-sph.github.io}}.
MagIC solves the 3D MHD equations in a spherical shell using a poloidal-toroidal decomposition for the velocity and the magnetic field,
\begin{align}
\vec{u} &= \vec{\nabla} \times \vec{\nabla} \times \left(W \  \vec{e_r}\right) + \vec{\nabla} \times \left(Z \  \vec{e_r}\right), \\
\vec{B} &= \vec{\nabla} \times \vec{\nabla} \times \left(b \  \vec{e_r}\right) + \vec{\nabla} \times \left(a_j \  \vec{e_r}\right),
\end{align}
where $W$ and $Z$ are respectively the poloidal and toroidal kinetic scalar potentials, while $b$ and $a_j$ are the magnetic ones.
The scalar potentials and the reduced pressure $p'$ are decomposed on spherical harmonics for the colatitude $\theta$ and the longitude $\phi$ angles, together with Chebyshev polynomials in the radial direction. The time stepping scheme is a mixed implicit-explicit scheme: A Crank-Nicolson scheme is used to advance the linear terms and a second order Adams-Bashforth scheme to advance the nonlinear terms and the Coriolis force. The linear terms are computed in the spectral space, while the nonlinear terms and the Coriolis force are computed in the physical space and transformed back to the spectral space. For more detailed descriptions of the numerical method and the associated spectral transforms, the reader is referred to \citet{1981ApJS...45..335G}, \citet{1997JFM...332..359T} and \citet{CHRISTENSEN2015245}.

All the simulations presented in this paper were performed using a standard grid resolution of $(n_r,n_{\theta},n_{\phi})=(257,512,1024)$.
The resolution was chosen to ensure that the dissipation scales are resolved. Indeed, the maxima of viscous and resistive dissipation are respectively at the spherical harmonic orders $l_{\nu}\simeq40$ and $l_{\eta}\simeq85$, which means that around ten cells resolve the resistive maximum.

\subsection{Initial conditions}\label{s:ic}

Many core-collapse simulations with a fast rotating progenitor have shown that the PNS is differentially rotating for several hundreds of milliseconds \citep[e.g,][]{2003ApJ...584..954A,2006ApJS..164..130O,2018JPhG...45h4001O,2020MNRAS.492...58B}.
The contraction of the PNS and accretion on the PNS accelerates it while the angular momentum is transported outside the PNS by the MRI.
However, it would not be possible to sustain a differentially rotating profile within an isolated shell of the PNS and in presence of strong magnetic fields, unless some forcing mechanism were to be applied at the boundaries. Therefore, in our simplified setup, we chose to artificially force the differential rotation by keeping the initial differential rotation profile at the outer boundary constant during the duration of the simulation.
This leads to a quasi-stationary state that is simpler to study and to compare among different models. The internal rotation profile of this quasi-stationary state can evolve and is determined by the combined effects of the backreaction of the magnetic field and the rotational forcing.

As initial and outer boundary condition, we used a cylindrical rotation profile inspired by core-collapse simulations with a central part that rotates like a solid body and an outer part that is differentially rotating with a power law dependency:
\begin{equation}
\Omega(s) = \frac{\Omega_\mathrm{i}}{\left(1 + \left(\frac{s}{0.4r_\mathrm{o}}\right)^{20q_0}\right)^{0.05}}
\,,
\end{equation}
where $s$ is the cylindrical radius, $q_0$ corresponds to the shear rate $q \equiv   - \frac{r}{\Omega}\frac{d\Omega}{dr}$ in the outer part, and $\Omega_\mathrm{i}$ is the rotation rate of the inner core. This rotation rate $\Omega_\mathrm{i}$ was computed so that the ratio of total angular momentum over moment of inertia is equal to the frame rotation rate $\Omega_0$ defined in Sect.~\ref{Eq_intro}.
  \begin{figure}
  \centering
   \resizebox{\hsize}{!}
      {
     \includegraphics[width=6cm]{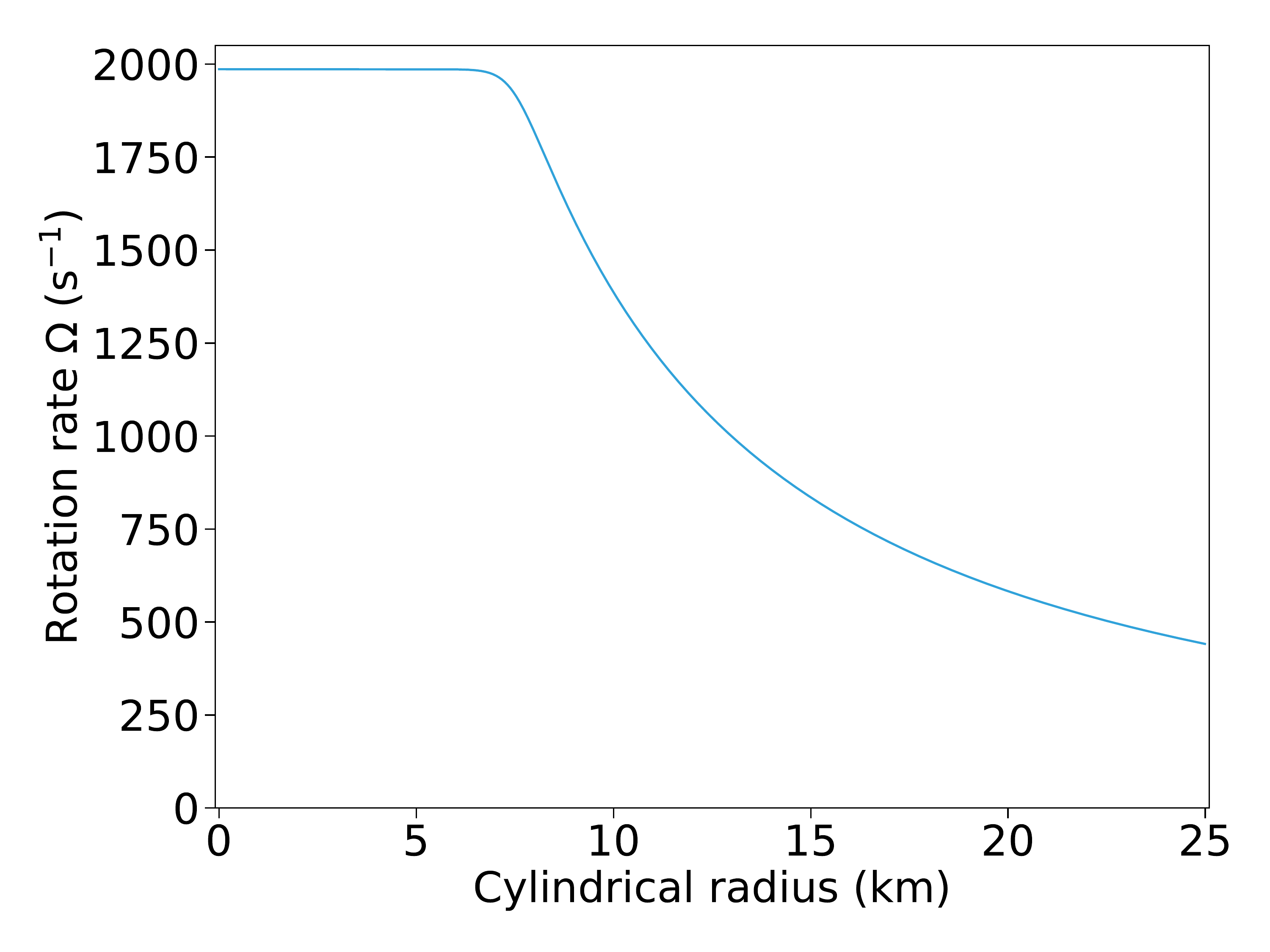}}
      \caption{Rotation profile inside the PNS as a function of the cylindrical radius. The inner region ($s<10$ km) is in solid body rotation and the outer region ($s>10$ km) follows a power law $\Omega \propto r^{-q_0}$ with $q_0=1.25$.}

         \label{rot_profile}
   \end{figure}
The profile shown in Fig. \ref{rot_profile} with $q_0=1.25$ has a smooth transition between solid body rotation when $s < 0.4r_\mathrm{o}$ and power law differential rotation when $s > 0.4r_\mathrm{o}$.

\begin{figure}
  \centering
  \includegraphics[width=6cm]{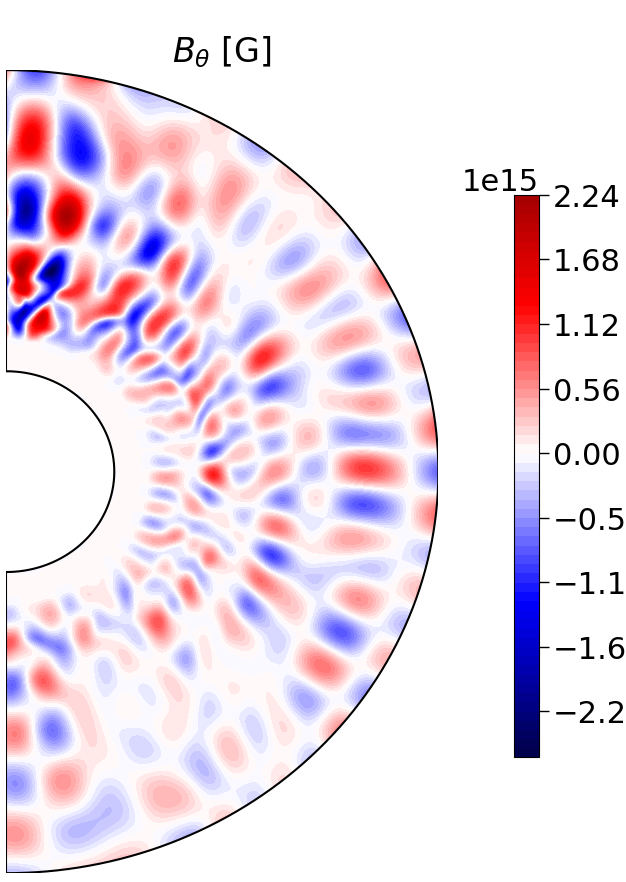}
  \caption{Slice of the $B_{\theta}$ component of the initial magnetic field for the meridional plane $\phi = 0$ with $[L_\mathrm{min},L_\mathrm{max}]=[0.23 r_\mathrm{o},0.38 r_\mathrm{o}]$ and $B_0=8.8\times 10^{14}\ \mathrm{G}$.}
  \label{Btheta_profile}
\end{figure}

Our model assumes a weak progenitor magnetic field and is designed to study in situ magnetic field amplification by the MRI, especially the large-scale field generation.
Several local studies have already shown an efficient amplification of the magnetic field on small scales \citep[e.g,][]{2009A&A...498..241O,2015ApJ...798L..22M,2015MNRAS.450.2153G,2016MNRAS.460.3316R}.
Therefore, the magnetic field may be initialized by small-scale modes with magnetic field strength high enough to resolve the MRI fastest growing mode.
The initial poloidal magnetic potential is a random superposition of modes with spherical harmonics indices $(l,m)$ and a radial profile of the form
\begin{equation}
b_{l,m} \propto  \left\{
 \begin{array}{ll}
 0 & \mbox{ for } r < 7.5 \mbox{ km} \\
\cos(k_r (r-r_\mathrm{i})) & \mbox{ for } r >12.5 \mbox{ km} \\
\end{array}
\right.
\,,
\end{equation}
where $k_r$ is the radial wavenumber and has a smooth transition between \SI{7.5}{\km} and \SI{12.5}{\km}.
We selected Fourier radial modes and spherical harmonic modes, whose wavelengths $\lambda_r = 2 \pi/k_r $ and $\lambda_l = \sqrt{ \frac{r^2}{l(l+1)}}$ lie within the range [$L_\mathrm{min},L_\mathrm{max}]$.
We note that this initial magnetic field has a vanishing net magnetic flux over the simulation domain.
The amplitude of these modes is initialized randomly with a wavelength dependence that provides a flat energy spectrum.
Figure \ref{Btheta_profile} shows an example of the initial poloidal magnetic field on a $\phi$-slice with a typical value of $[L_\mathrm{min},L_\mathrm{max}]=[0.23 r_\mathrm{o},0.38 r_\mathrm{o}]$.

The initial root mean square magnetic field strength $B_0$ is varied from
$B_0 = \SI{6.31e14}{G} $ to $B_0 = \SI{3.36e15}{G}$.
The fastest growing mode of MRI in ideal MHD \citep{1991ApJ...376..214B} has a wavelength of
\begin{equation}
\lambda_\mathrm{MRI} = \frac{8\pi}{q_0(4-q_0)\Omega_0}\frac{B_0}{\sqrt{4\pi\rho_0}}
\,.
\end{equation}
It ranges from $\lambda_\mathrm{MRI} =\SI{1.9e5}{\cm}$ to $\lambda_\mathrm{MRI} = \SI{1.0e6}{\cm}$ and is well resolved with our resolution of $\Delta_r \equiv {D}/{N_r}=\SI{7.32e3}{\cm}$.

\subsection{Boundary conditions}

We assumed non-penetrating boundary conditions ($u_r=0$). At the inner boundary, we used a standard no-slip condition with an inner core in solid body rotation at a rate $\Omega_\mathrm{i}$ that evolves with the viscous torque (i.e., $u_{\phi}=\Omega_\mathrm{i} r_\mathrm{i}$ and $u_{\theta}=0$). At the outer boundary, we used a modified no-slip condition where we forced $u_{\phi}$ to match the initial rotation profile at all times (and $u_{\theta}=0$). The outer boundary is therefore not in solid body rotation.
For the magnetic field, we compared three different boundary conditions: pseudo-vacuum (imposing a radial field: $\vec{B} \times \vec{n} =0$, where $\vec{n}$ is the normal vector of the outer surface), perfect conductor (imposing a tangential field: $\vec{B}\cdot \vec{n} = 0$), and insulating (matching a potential field outside the domain).

\subsection{Physical parameters and dimensionless numbers}

We chose the physical parameters of the simulations to represent a fast rotating PNS model similar to the study of \citet{2015MNRAS.447.3992G}.
All of our models have a uniform viscosity of $\nu = \SI{7.03e11} {\cm \square \per \s}$.
Given the values for the viscosity, the rotation rate $\Omega_0$ and the characteristic length $D$ (Sect. \ref{Eq_intro}), the dimensionless Ekman number (characterizing the importance of viscosity over Coriolis force) is
\begin{equation}
E \equiv \frac{\nu}{\Omega_0 \ D^2} = 2 \times 10^{-4}.
\end{equation}
For a PNS, the viscosity is large due to the impact of neutrinos while the resistivity is very small, which leads to a high magnetic Prandtl number $P_\mathrm{m} \equiv \nu/\eta$. A realistic value of the magnetic Prandtl number for a PNS is $P_\mathrm{m} \approx 10^{13}$ according to \citet{1993ApJ...408..194T} and \citet{2007ApJ...655..447M}.
However, numerical simulations for these values of $P_\mathrm{m} $ are not possible due to numerical constraints.
We varied the resistivity from $\eta = \SI{4.39e10}{\cm\square\per\s}$ to $\eta = \SI{1.66e11}{\cm\square\per\s}$, which corresponds to magnetic Prandtl numbers in the range
$P_\mathrm{m} \in \left[ 4, \ 8,\ 16\right]$.
Our standard value $P_\mathrm{m} = 16$ is a compromise between the wish to be in the high magnetic Prandtl regime and the computing time constraints.
The corresponding magnetic Reynolds number that characterizes the relative importance of magnetic advection to magnetic diffusion (resistivity) is
\begin{equation}
R_\mathrm{m} \equiv \frac{D^2 \ \Omega_0}{\eta} \in \left[ 2 \times 10^{4}, \ 4 \times 10^{4}, \ 8 \times 10^{4}\right].
\end{equation}

\section{Results}
\label{results}

\subsection{Typical quasi-stationary dynamos}
\label{MRI_example}

We first describe the results from one fiducial simulation where we obtain a self-sustained dynamo.
For the model \texttt{Standard} (see Table~\ref{table:annex}), we applied insulating boundary conditions and initialized the magnetic field with the parameters $B_0=8.8\times 10^{14}\ \mathrm{G}$ and $[L_\mathrm{min},L_\mathrm{max}] = [0.23 r_\mathrm{o}, 0.38 r_\mathrm{o}]$ (see Sect.~\ref{s:ic}).

\begin{figure}
  \centering
   \resizebox{\hsize}{!}
            {
  \includegraphics[width=6cm]{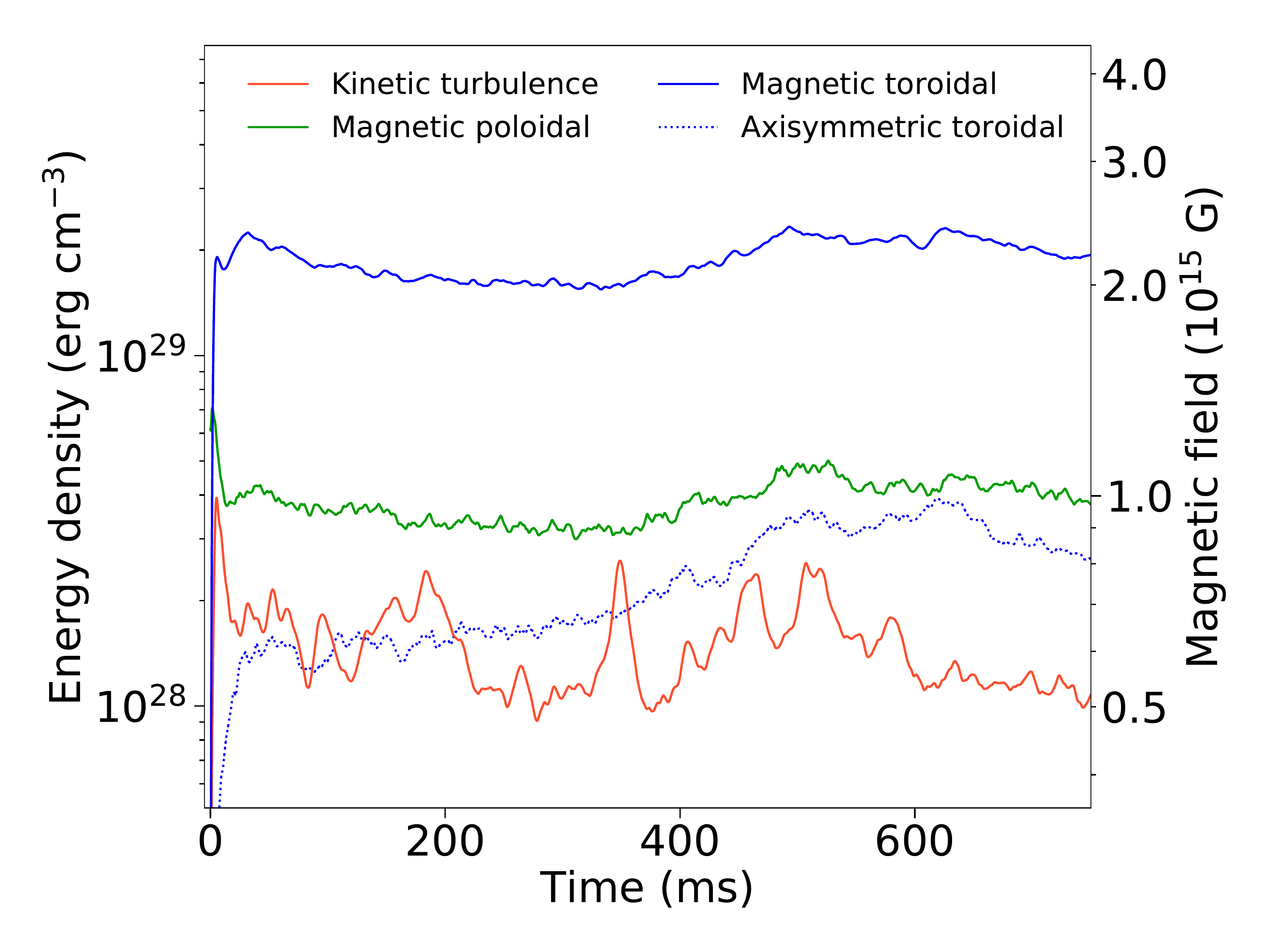}}
  \caption{Temporal evolution of the magnetic and turbulent kinetic energy density for the model \texttt{Standard} $B_0=8.8\times 10^{14}\ \mathrm{G}$, $[L_\mathrm{min},L_\mathrm{max}] = [0.23 r_\mathrm{o}, 0.38 r_\mathrm{o}]$. The blue and green lines are the toroidal and poloidal contributions of the magnetic energy density, while the orange line is the turbulent kinetic energy density (axisymmetric toroidal contribution is removed). The blue dotted line is the axisymmetric contribution to the toroidal magnetic energy density.}
  \label{Ts_energies}
\end{figure}

Figure \ref{Ts_energies} shows the temporal evolution of the volume-averaged toroidal and poloidal magnetic energy densities and the turbulent kinetic energy density.
In order to separate the MRI-driven turbulent flow from the differential rotation, the turbulent kinetic energy density is computed by subtracting the contribution of the axisymmetric azimuthal velocity from the averaged kinetic energy density. After approximately 400 milliseconds, we obtain a statistically stationary state with a mean magnetic field intensity $B = 2.5 \times 10^{15}\ \mathrm{G}$. The main contribution is from the toroidal magnetic field, which is $\sim$2 times larger than the poloidal magnetic field.
The magnetic field is predominantly nonaxisymmetric, the axisymmetric magnetic field being $\sim$3~times weaker than the total magnetic field.
The averaged magnetic energy density is more than ten times stronger than the turbulent kinetic energy density. A similar ratio is observed in local models of MRI-driven turbulence (see Sect. \ref{Comp}).
   \begin{figure}
     \centering
     \includegraphics[width=0.5\textwidth]{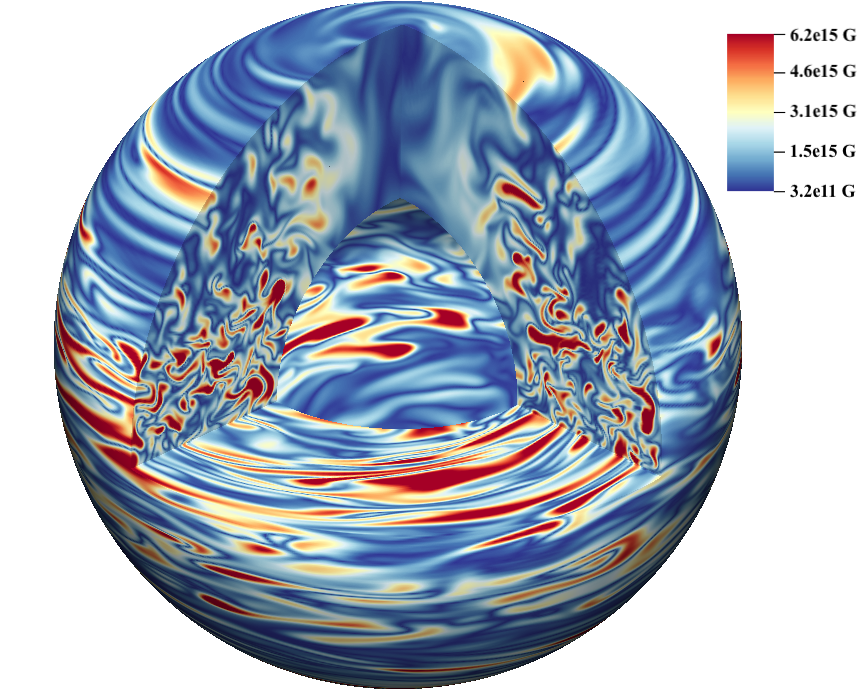}
     \centering
     \includegraphics[width=0.5\textwidth]{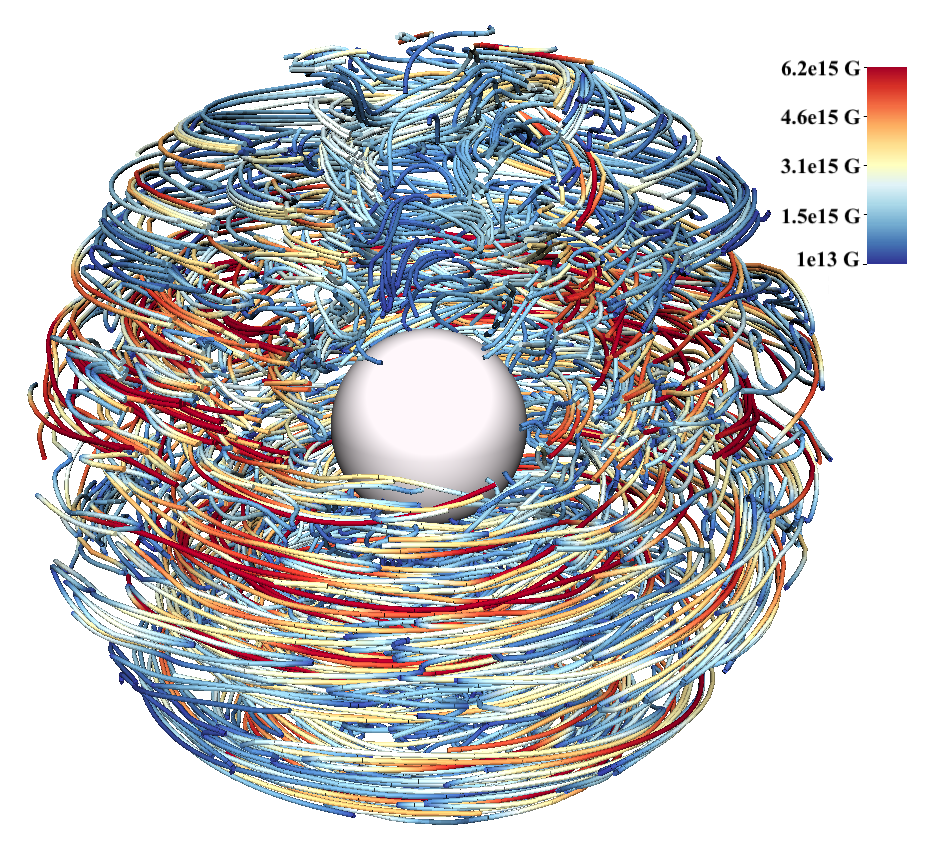}
      \caption{Top: 3D snapshot of the intensity of the magnetic field at $t=\SI{600}{\ms}$ for model \texttt{Standard}. The colors represent the magnetic field amplitude from weak (blue) to strong (red). 
      Bottom: 3D snapshot of the magnetic field lines at $t=\SI{600}{\ms}$ for the model \texttt{Standard}. The colors represent the magnetic field amplitude.}

         \label{3D_figure}
   \end{figure}
\begin{figure}
  \centering
  \includegraphics[width=7cm]{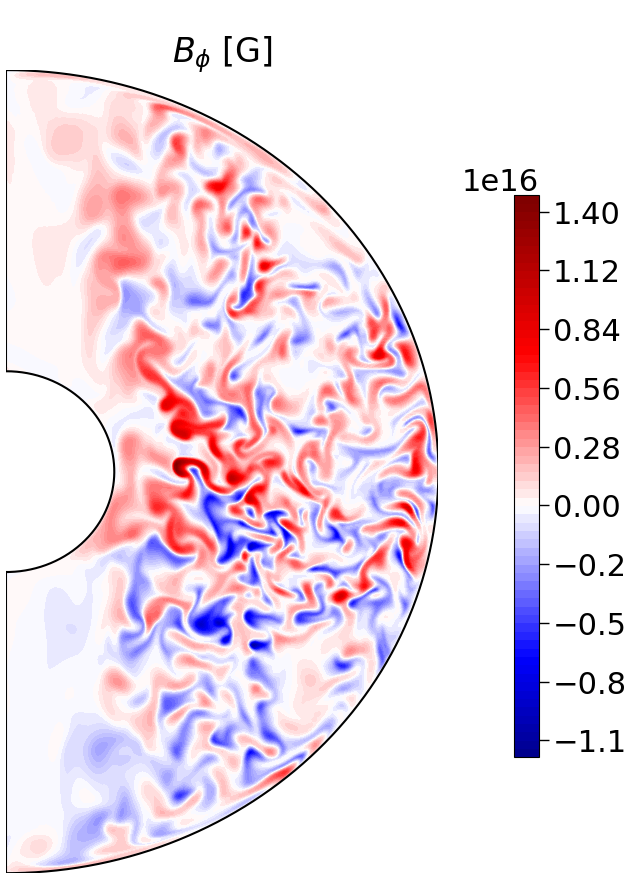}
  \includegraphics[width=7cm]{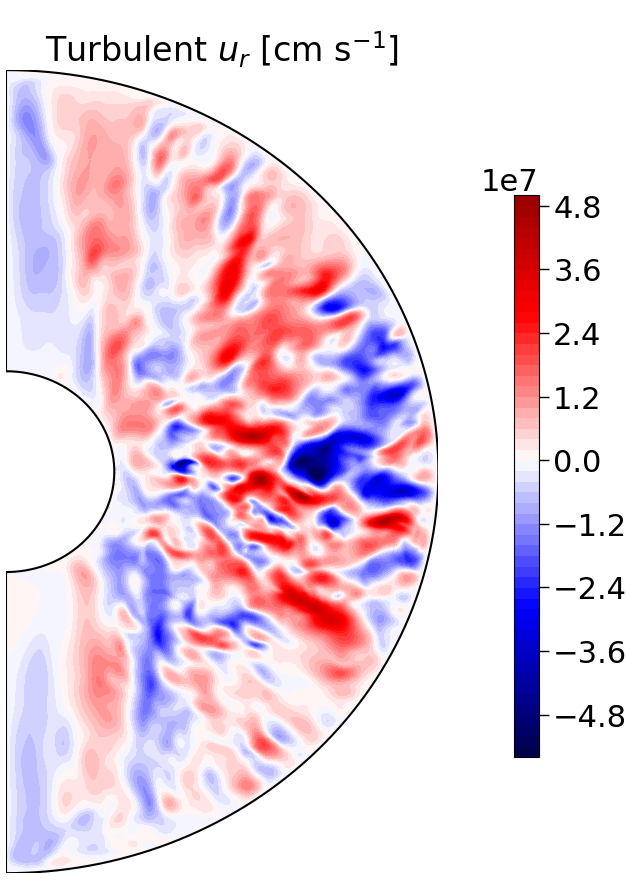}
\caption{Slices for $\phi = 0$ and $t=600$ ms for the model \texttt{Standard}. Top: Azimuthal magnetic field $B_{\phi}$. Bottom: Turbulent radial velocity $u_r$ (i.e., the axisymmetric contribution is subtracted from the radial velocity).}
  \label{Bphi_snap}
\end{figure}

\begin{figure}
  \centering
   \resizebox{\hsize}{!}
            {
  \includegraphics[width=6cm]{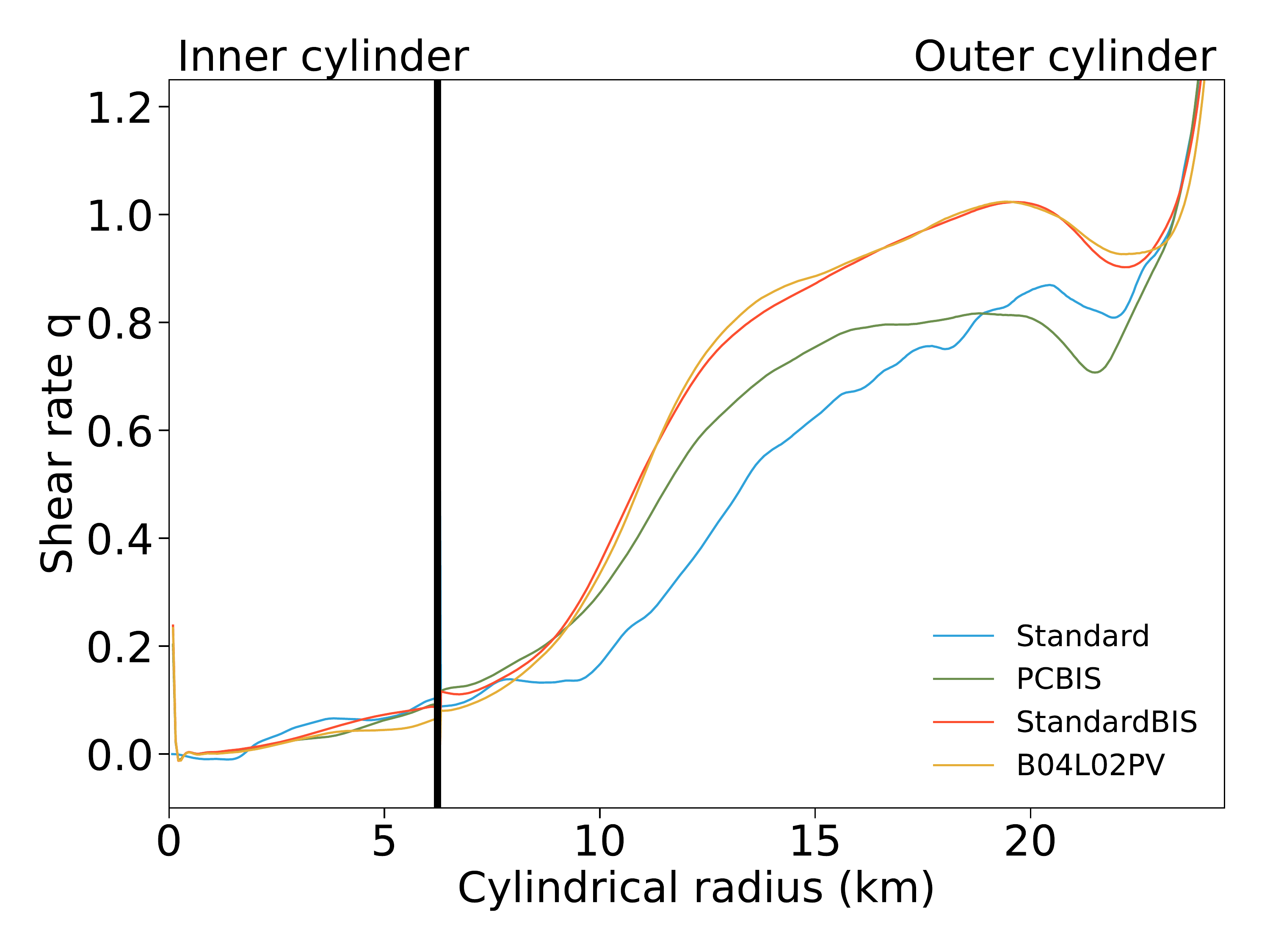}}
  \caption{Comparison of the vertically and azimuthally averaged shear rates $q$ inside the PNS for different MHD simulations as a function of the cylindrical radius. The values are time-averaged in the quasi-stationary phase of the dynamo. The black line represents the radius of the inner core $r_i$.}
  \label{shear_profile}
\end{figure}

Figures \ref{3D_figure} and \ref{Bphi_snap} are representative snapshots of the quasi-stationary state and illustrate the complex geometry of the magnetic field due to MRI-driven turbulence.
As is common in this turbulence, the winding of the magnetic field by the shear produces elongated structures in the azimuthal direction, which are clearly seen in the equatorial plane (Fig. \ref{3D_figure}).
On the meridional cuts, the magnetic field is distributed on smaller scales.
The turbulent structure of the magnetic field has lost memory of the initial magnetic configuration.
Near the rotation axis, the magnetic field is not very intense and the MRI-driven turbulence is weaker for cylindrical radii smaller than $s_\mathrm{turb} \approx \SI{9.4}{\km}$.
In the most turbulent zone, the turbulent radial velocity field has similar small-scale structures, while weak large-scale flows develop near the rotation axis (bottom panel of Fig. \ref{Bphi_snap}).
The weak magnetic field and the non-turbulent flow near the rotation axis are expected because the vertically averaged shear rate of the simulation is small for $s\le $ \SI{9}{\km} and the inner part of the radial profile rotates as a solid body (blue line in Fig. \ref{shear_profile}).

\begin{figure}
\centering
\resizebox{\hsize}{!}
{
\includegraphics{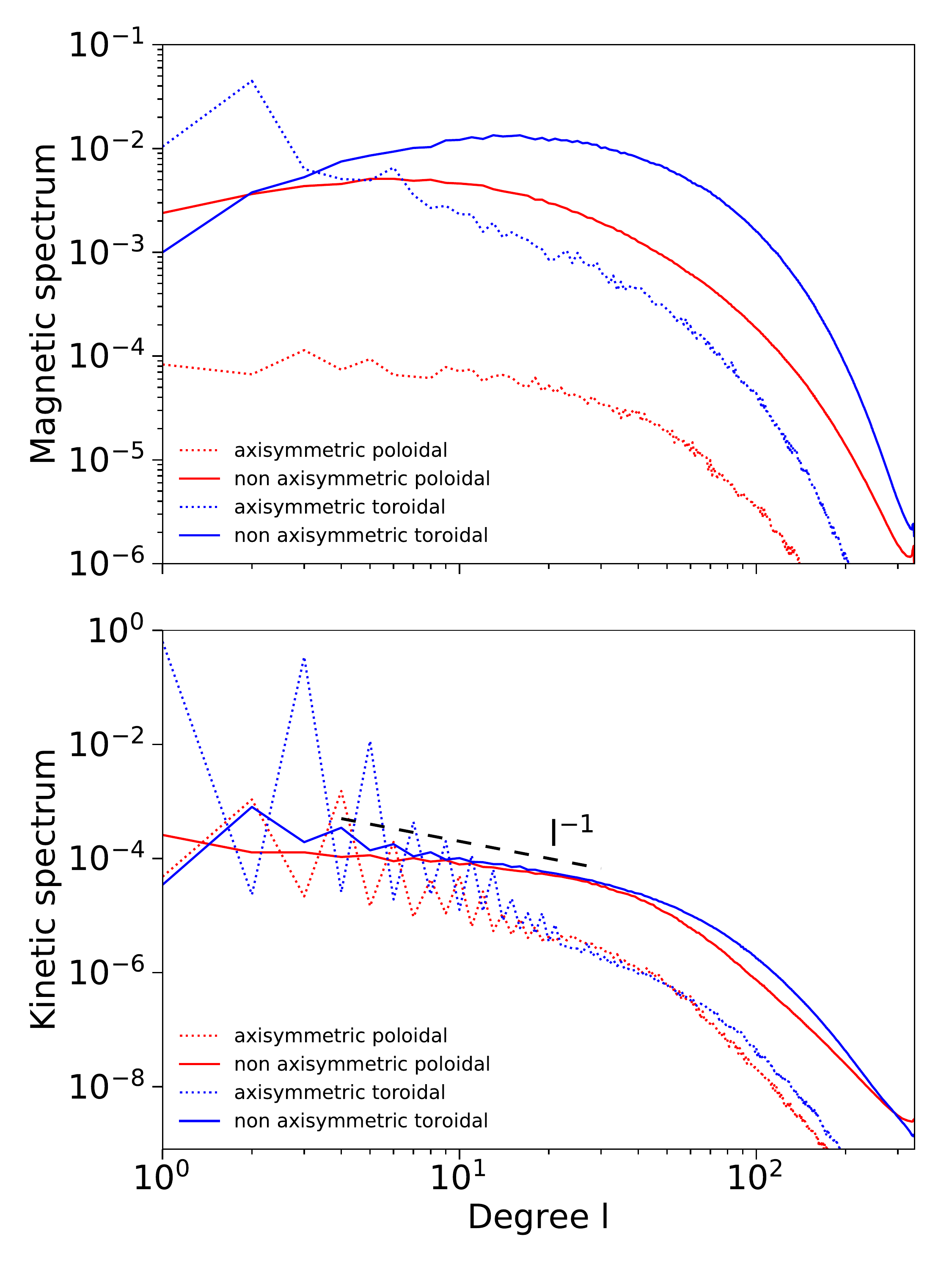}}
\caption{Top: Spectrum of the toroidal magnetic energy (blue) and the poloidal magnetic energy (red) normalized by the total magnetic energy as a function of the spherical harmonics order $l$. The dotted lines corresponds to the axisymmetric components of these energies. Bottom: Spectrum of the toroidal kinetic energy (blue) and the poloidal kinetic energy (red) normalized by the total kinetic energy as a function of the spherical harmonics order $l$. The dotted lines corresponds to the axisymmetric components. The black dashed line corresponds to a scaling of $l^{-1}$.}
\label{standard_spectrum}
\end{figure}

To understand how the magnetic and kinetic energies are distributed over different scales, we computed the axisymmetric and nonaxisymmetric components of the toroidal and poloidal spectra (Fig. \ref{standard_spectrum}).
To first order, the nonaxisymmetric  toroidal magnetic spectrum as a function of the spherical harmonics order $l$ (top panel of Fig. \ref{standard_spectrum}) can be decomposed in two parts: an increasing part up to order $l\simeq 20$ and a sharp decrease for the small scales where the dissipation occurs.
The axisymmetric toroidal contribution dominates for the largest scales at $l<3$, while it is clearly subdominant for the small scales. The quadrupole mode ($l=2$) is particularly strong for this model but this varies between different models.
Contrary to the toroidal component, the poloidal magnetic field is dominated by its nonaxisymmetric component at all scales.
The nonaxisymmetric poloidal magnetic spectrum is similar to the toroidal one but peaks at larger scales for $l \simeq 10$.
These spectra show that the main contribution to the total magnetic energy comes from the toroidal component at intermediate scales ($l\sim 40$).

The nonaxisymmetric poloidal and toroidal kinetic spectra are similar (bottom panel of Fig. \ref{standard_spectrum}) and can also be decomposed in two parts: a power law for the large scales and a sharp decrease for the small scales. The power law at larger scales ($l<35$) seems to match a scaling of $l^{-1}$, which is consistent with the kinetic spectrum of high Reynolds local simulations \citep{2010A&A...514L...5F}.
The axisymmetric toroidal component of the kinetic spectra is composed of the turbulence contribution dominating at small scales and the differential rotation dominating at large scales.
The oscillations between odd and even modes observed for $l<20$ are due to the symmetry of the differential rotation with respect to the equatorial plane.
In the same way, the axisymmetric poloidal contribution contains both turbulence (for $l>20$) and the meridional circulation (for $l<20$), which originates from the interplay between angular momentum transport and the outer boundary forcing.
The total kinetic spectrum is dominated by the axisymmetric differential rotation and the meridional circulation at the large scales ($l<10$), while the nonaxisymmetric kinetic turbulence dominates at the intermediate and small scales.

\begin{figure}
  \centering
  \resizebox{\hsize}{!}
            {
     \includegraphics[width=6cm]{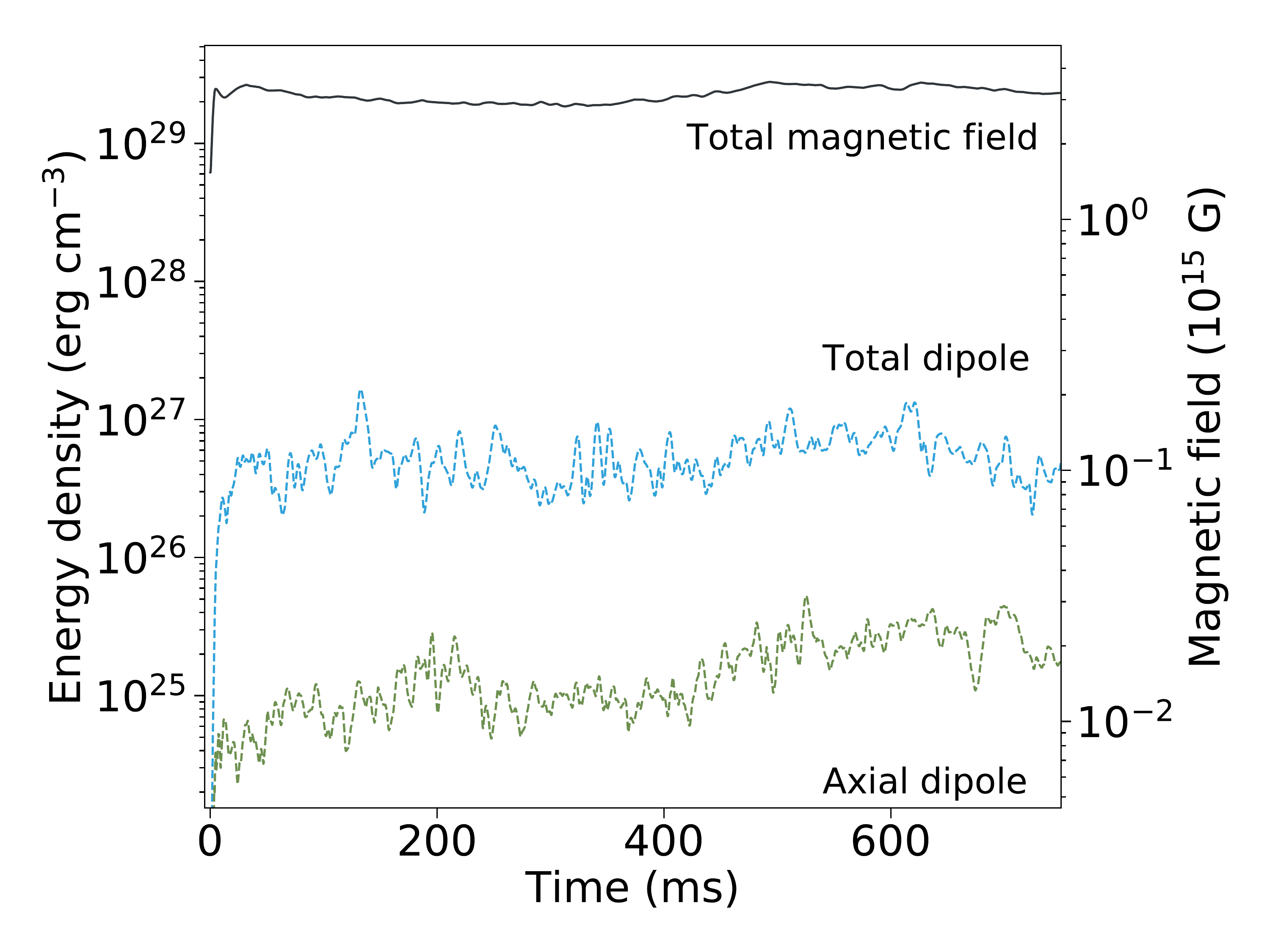}}
      \caption{Temporal evolution of the averaged magnetic energy density and dipolar energy density for the model \texttt{Standard}. The black line is the averaged magnetic energy density, the blue dash-dotted line is the averaged dipole energy density and the green dash-dotted is the axial dipole energy density.}
         \label{Ts_dip}
   \end{figure}

Since the magnetar timing parameters only constrain the dipolar component of the magnetic field, we focus in more details on this specific mode.
The dipole field strength reaches a significant intensity of $B_\mathrm{dip} = 1.25 \times 10^{14}\ \mathrm{G}$ but it is not the dominant mode in the simulation as it is approximately 20 times weaker than the total magnetic field intensity (see blue and black lines in Fig. \ref{Ts_dip}). This is consistent with a visual inspection of the snapshots, where a dipole structure of the magnetic field cannot be seen.

\begin{figure}
  \centering
   \resizebox{\hsize}{!}
            {
     \includegraphics[width=6cm]{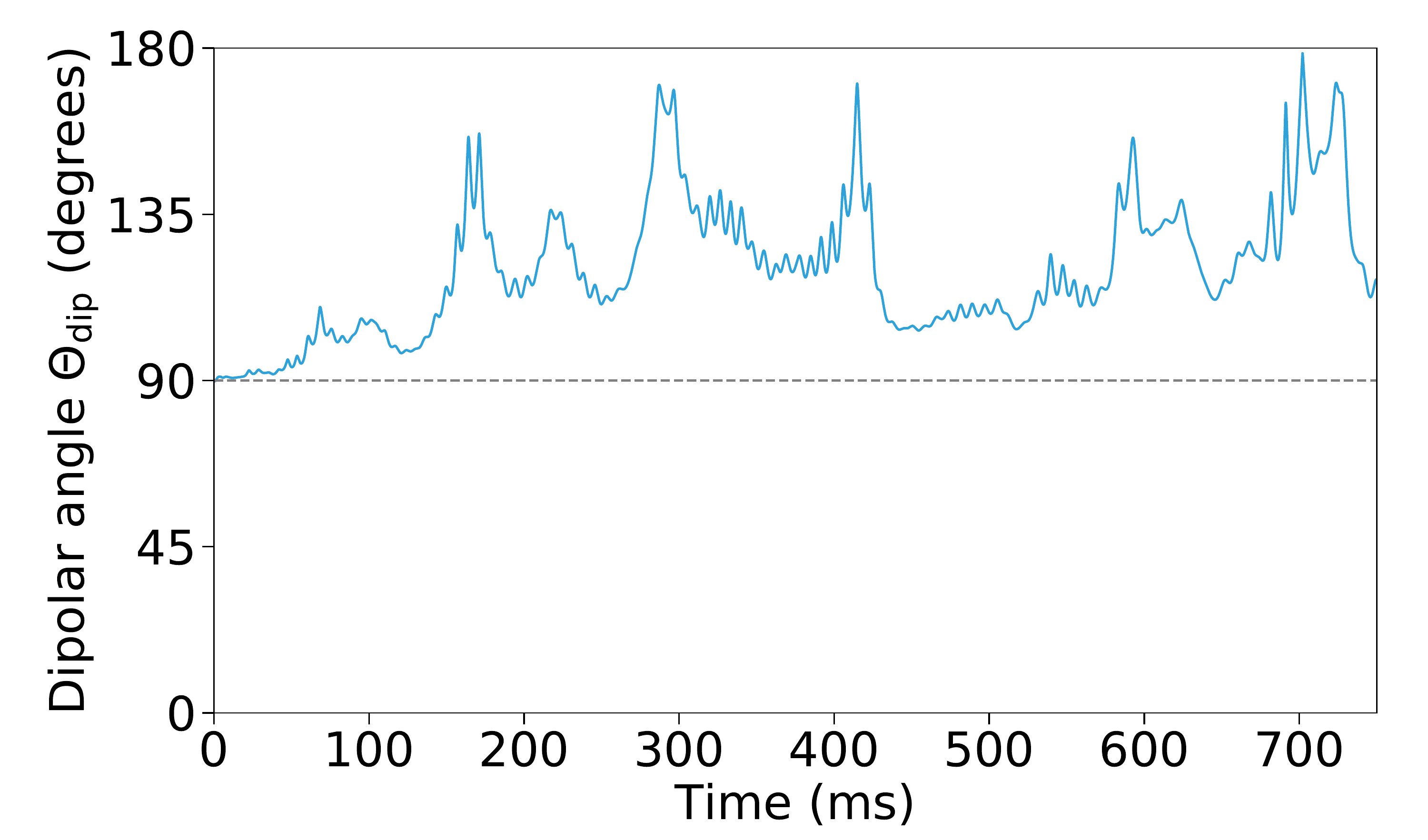}}
      \caption{Temporal evolution of the dipolar tilt angle in degrees for the model \texttt{Standard}. The aligned dipole corresponds to $\theta_\mathrm{dip} = 0^{\circ}$ and the equatorial dipole corresponds to $\theta_\mathrm{dip} = 90^{\circ}$.}
         \label{Dipole_angle}
   \end{figure}

In Fig. \ref{Ts_dip}, the intensity of the axial dipole is $\sim$4 times lower than the averaged dipole intensity, implying that the dipole is tilted toward the equatorial plane.
We computed the dipole tilt angle~$\theta_\mathrm{dip}$ from the magnetic dipole moment $\vec{\mu}$\
given by the formulas
\begin{equation}
\vec{\mu}= \frac{1}{2}\iiint \vec{r} \times \vec{J} dV,
\end{equation}
\begin{equation}
\theta_\mathrm{dip} = 90^{\circ} -  \arctan\left(\frac{\mu_z}{\sqrt{\mu_x^2+\mu_y^2}}\right),
\end{equation}
where $\vec{J} = \nabla \times \vec{B}/\mu_0$ is the current and the volume integral covers the entire numerical domain. Figure~\ref{Dipole_angle} shows the time series of the dipole tilt angle.
Its time-averaged value is $\theta_\mathrm{dip} = 120^{\circ}$ for the run \texttt{Standard}. This result is consistent with the ratio of the energy contained in the total dipole and axial dipole.
The equatorial character of the dipole may be explained in the following way: The azimuthal magnetic field does not contribute to the axial dipole but it can contribute to the equatorial dipole through its $m=1$ component. The fact that the MRI generates a predominantly azimuthal magnetic field may therefore explain the dominance of the equatorial dipole.
The peaks we see in the dipole tilt angle evolution are due to the weakening of the dipole moment in the equatorial plane for short periods of time.

\subsection{Dynamo threshold}
\label{Threshold}

We showed in the previous subsection that a MRI-driven dynamo can reach a quasi-stationary state but the turbulence can also be damped by the diffusion processes depending on the diffusivities or the initial magnetic field strength.
In fact, diffusion processes (viscosity and resistivity) tend to limit the growth of MRI modes for weak fields \citep{2007A&A...476.1123F,2015MNRAS.447.3992G}.
Therefore, we expect the dynamo threshold to depend on the initial magnetic field intensity $B_0$ and the magnetic Prandtl number $P_\mathrm{m}$ (in this study, the Ekman number $E$ is the same for all simulations).

\begin{figure}
  \centering
   \resizebox{\hsize}{!}
            {
  \includegraphics[width=6cm]{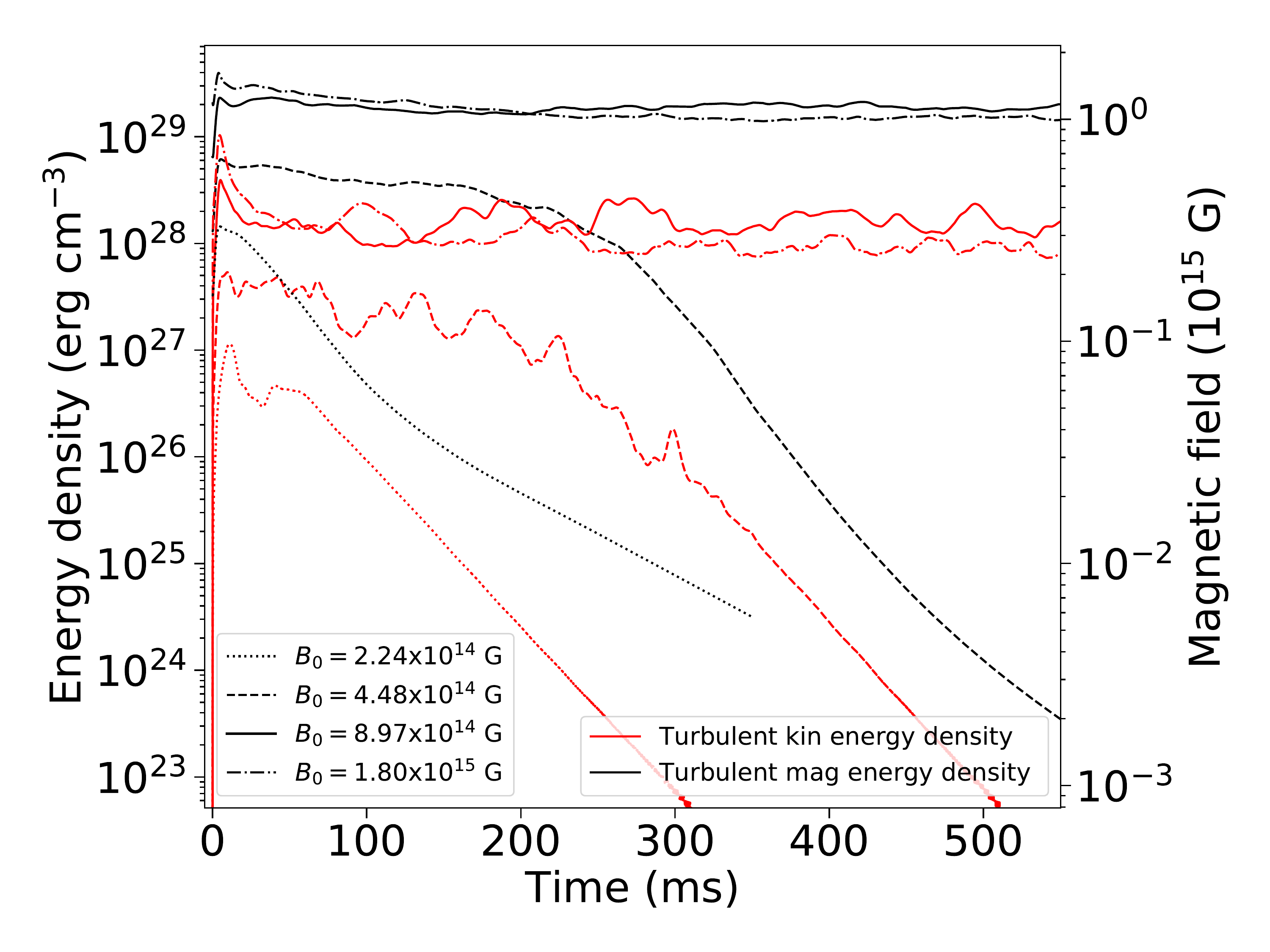}}
  \caption{Temporal evolution of the magnetic (black) and kinetic (red) turbulent energy densities for different values of $B_0$ with $[L_\mathrm{min},L_\mathrm{max}] = [0.23 r_\mathrm{o},0.38 r_\mathrm{o}]$ and $P_\mathrm{m}=16$.}
  \label{Ts_B_0}
\end{figure}

Figure~\ref{Ts_B_0} shows the time evolution of the magnetic and turbulent kinetic energies for different values of $B_0$. Three behaviors are observed:
For the lowest $B_0$ (dotted line), the turbulent energies decrease sharply with time, which indicates that no dynamo is achieved.
For intermediate $B_0$ (dashed line), the turbulent energies have a slow decreasing phase followed by a fast drop. This behavior is called a transient dynamo in the following sections.
For higher $B_0$, the turbulent energies reach a quasi-steady state, with some fluctuations around a non-evolving average.
In this case, the simulations reach a self-sustained quasi-stationary dynamo.

\begin{figure}
  \centering
   \resizebox{\hsize}{!}
            {
  \includegraphics[width=6cm]{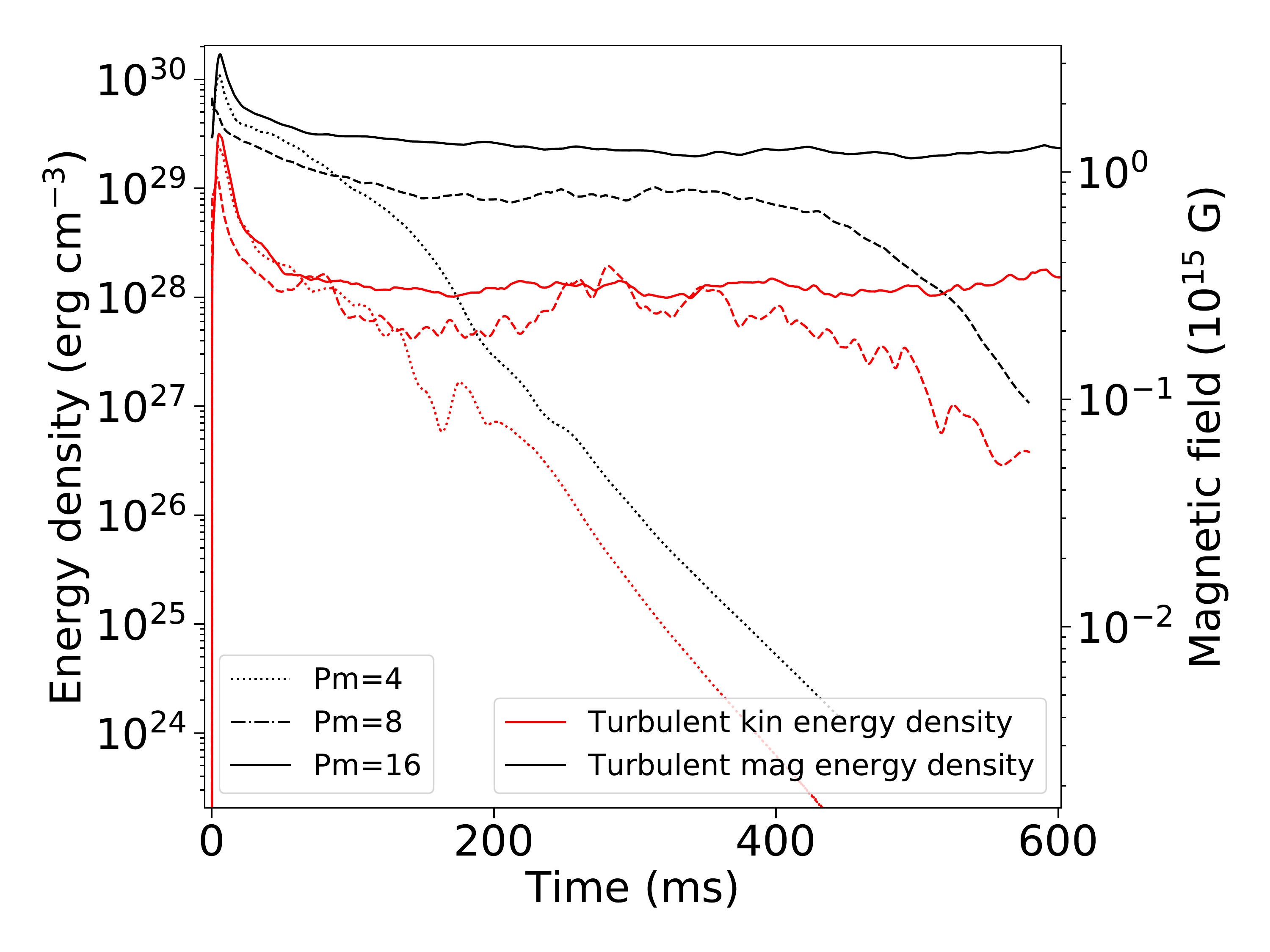}}
  \caption{Same as Fig. \ref{Ts_B_0} for three different values of $P_\mathrm{m}$ and with $[L_\mathrm{min},L_\mathrm{max}] = [0.75,1]$ and $B_0=1.8 \times 10^{15}$ G.}
  \label{Ts_Pm}
\end{figure}

Figure \ref{Ts_Pm} shows that a self-sustained dynamo can be obtained for $P_\mathrm{m}\gtrsim 12 \pm 4$.
A simulation with $P_\mathrm{m} = 16$ gives a quasi-stationary dynamo, $P_\mathrm{m} =8$ a transient dynamo, and $P_\mathrm{m}=4$ no dynamo.
This result is independent of the value of $B_0$ for $B_0 \ge \SI{e15} {G}$.
It should be noted that the threshold of the dynamo action is set for relatively strong initial magnetic fields because of the high values of viscous and magnetic diffusivities. For lower diffusivities relevant to a PNS, we would expect MRI dynamo to set in for lower initial magnetic fields.

\subsection{Robustness of the results} \label{Robustness}

The previous subsections demonstrated the ability of self-sustained MRI-driven dynamos to generate a significant, though subdominant, magnetic dipole. In this section, we assess the robustness of these results by studying the impact of initial and boundary conditions of the magnetic field.
We show that the quasi-stationary state is independent of the initial conditions and that the boundary conditions only have a minor impact.

\begin{figure}
  \centering
   \resizebox{\hsize}{!}
            {
      \includegraphics[width=6cm]{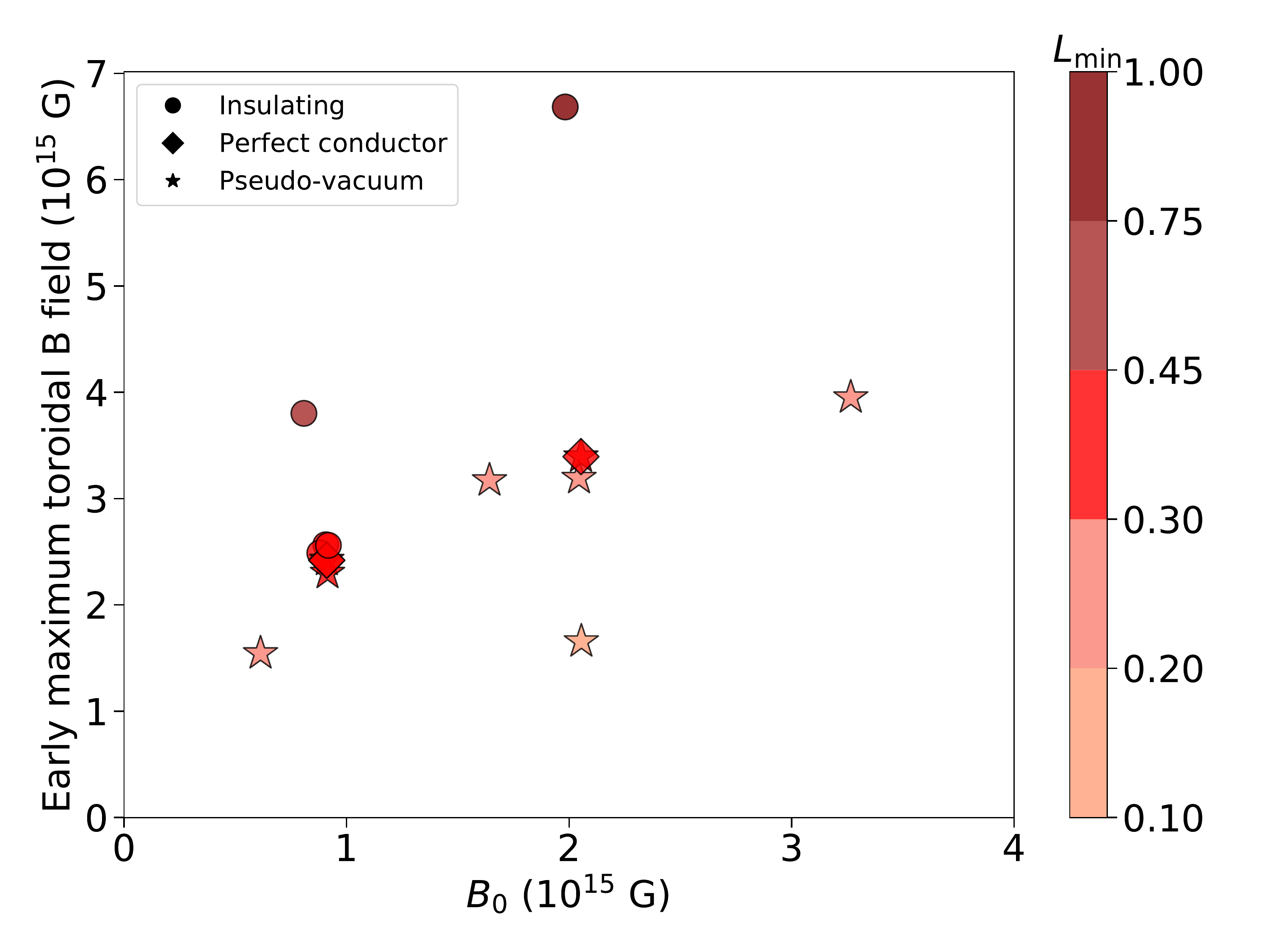}}
      \caption{Early maximum toroidal magnetic field as a function of initial magnetic field strength $B_0$. The symbol color describes the smallest initial magnetic length scale $L_\mathrm{min}$. The symbol shape represents the magnetic boundary conditions, between perfect conductor (diamond), insulating (circle) and pseudo-vacuum (star).}
         \label{max_amp_B0}
   \end{figure}

We varied the initial magnetic field amplitude $B_0$ and its minimum length scale~$L_\mathrm{min}$ and display the results in Fig. \ref{max_amp_B0}.
To characterize the early phase of magnetic amplification, we measured the early maximum of the toroidal magnetic field after a few milliseconds (i.e., the first local maximum value reached in the time series, see for example Fig.~\ref{Ts_Pm}).
For a given smallest initial magnetic length scale $L_\mathrm{min}$ (indicated by the symbol color), the early maximum toroidal magnetic field is clearly increasing with the initial magnetic amplitude by a factor of $\sim$2.
This dependence is shallower than in \citet{2016MNRAS.460.3316R}, probably because our initial magnetic field is on smaller scales.
On the other hand, for a given $B_0$, the early maximum toroidal magnetic field increases by a factor of $\sim$5 when $L_\mathrm{min}$ is increased.
The parameter~$L_\mathrm{min}$ seems to have the most prominent effect on the early maximum toroidal magnetic field.
A possible interpretation is that a small-scale magnetic field is more prone to dissipation.
Finally, Fig. \ref{max_amp_B0} suggests that the same initial conditions give approximately the same early maximum magnetic field, independently of the magnetic boundary conditions. Since this early maximum toroidal magnetic field is taken at the beginning of the simulation, boundary conditions may have a minor effect at this stage.

\begin{figure}
  \centering
   \resizebox{\hsize}{!}
            {
     \includegraphics[width=6cm]{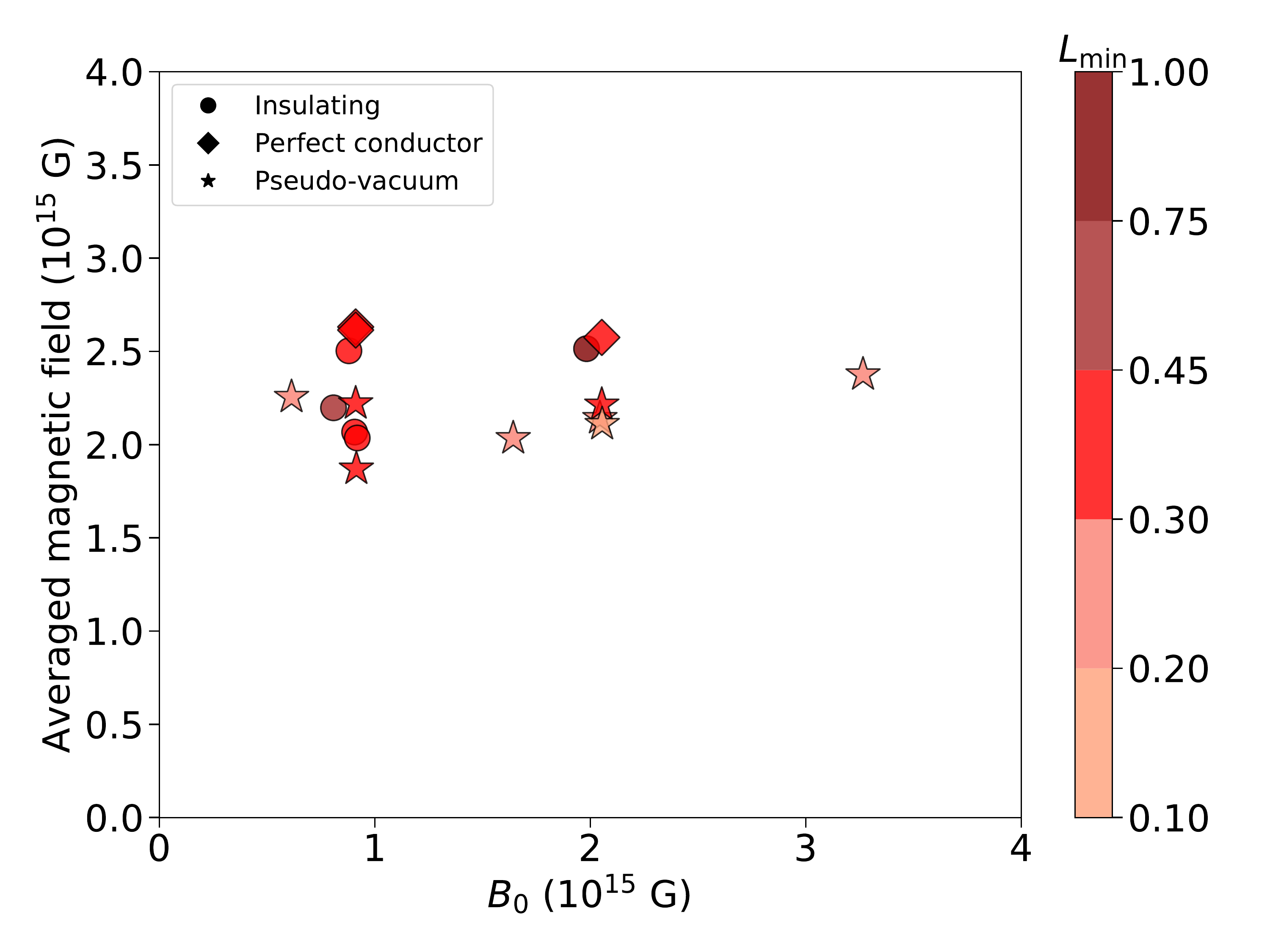}}
      \caption{Magnetic field strength averaged in the quasi-stationary state as a function of initial magnetic field strength~$B_0$. The symbol color describes the smallest initial magnetic length scale $L_\mathrm{min}$. The symbol shape represents the magnetic boundary conditions. This figure shows that the quasi-stationary state is roughly independent of the initial and boundary conditions, contrary to the early maximum.}
         \label{Btot_max_amp}
   \end{figure}

To characterize our self-sustained dynamos, we performed time and volume averages on the magnetic field once the simulation has reached a quasi-stationary state (Fig. \ref{Btot_max_amp}).
The averaged magnetic field now displays much less variation than the early maximum.
Indeed, all models lie between $1.8 \times 10^{15} \ \mathrm{G}$  and $2.7\times 10^{15}\ \mathrm{G}$, with a mean of $B = (2.27 \pm 0.23 )\times 10^{15} \ \mathrm{G}$.
Contrary to Fig. \ref{max_amp_B0}, no systematic impact of the initial conditions can be observed as different simulations give similar magnetic fields.
For a given boundary condition, the small variations can be explained by the stochasticity of MRI-driven dynamos.
On the other hand, the magnetic field is slightly stronger in the case of perfect conductor boundary conditions.
This trend may be explained by a stronger toroidal magnetic field close to the outer boundary, since the perfect conductor condition allows for a non-vanishing toroidal magnetic field.
Overall, these results indicate that the initial conditions and boundary conditions have a small impact on the global properties of the final quasi-stationary state.

\begin{figure}
  \centering
   \resizebox{\hsize}{!}
            {
     \includegraphics[width=6cm]{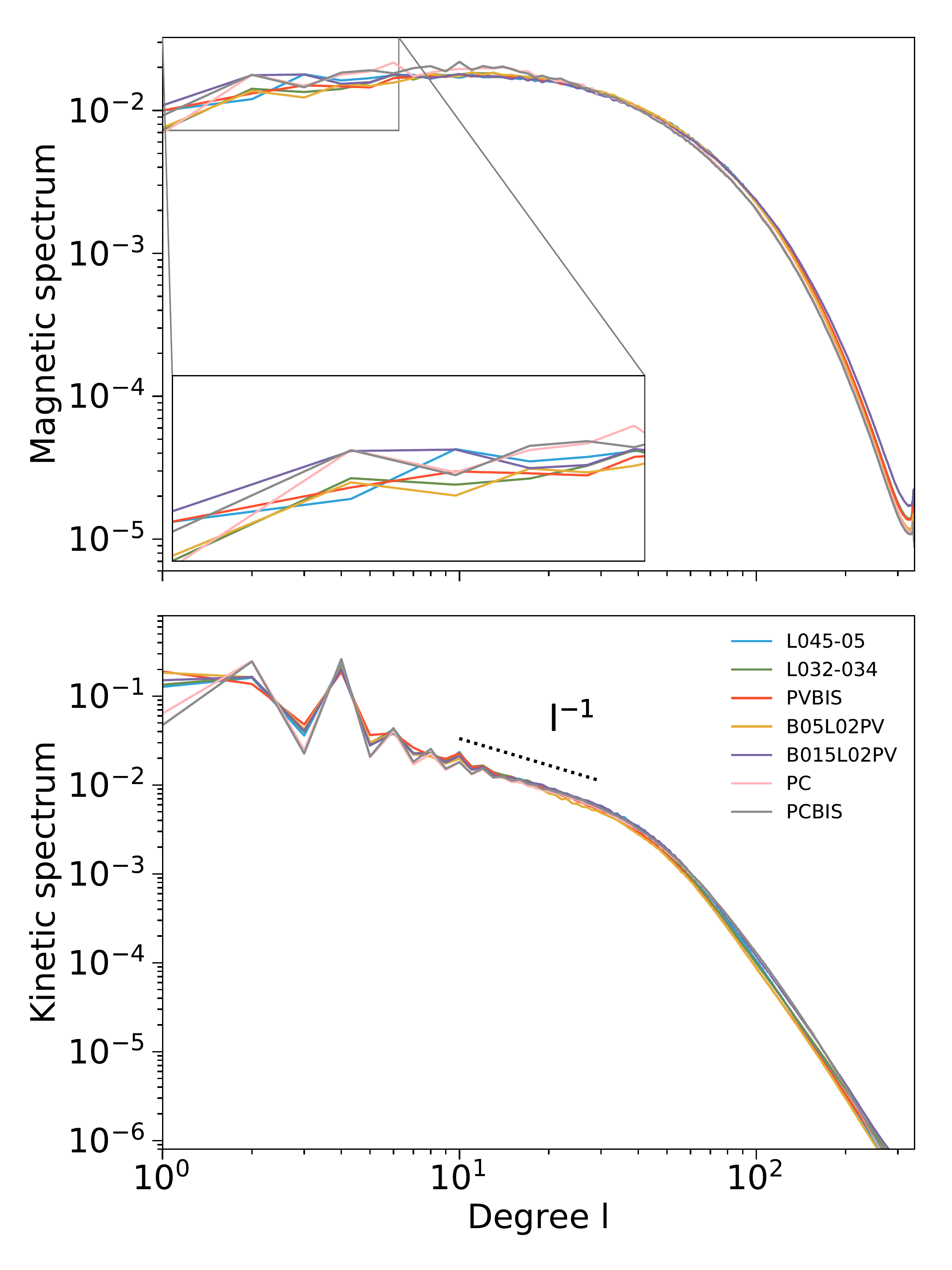}}
      \caption{Top: Spectrum of the magnetic energy as a function of the spherical harmonics order $l$. Bottom: Spectrum of the poloidal kinetic energy as a function of the spherical harmonics order $l$. Each line is from a different model. The dotted line corresponds to a scaling of $l^{-1}$.}
         \label{Spec_l}
   \end{figure}

\begin{figure}
  \centering
   \resizebox{\hsize}{!}
            {
     \includegraphics[width=6cm]{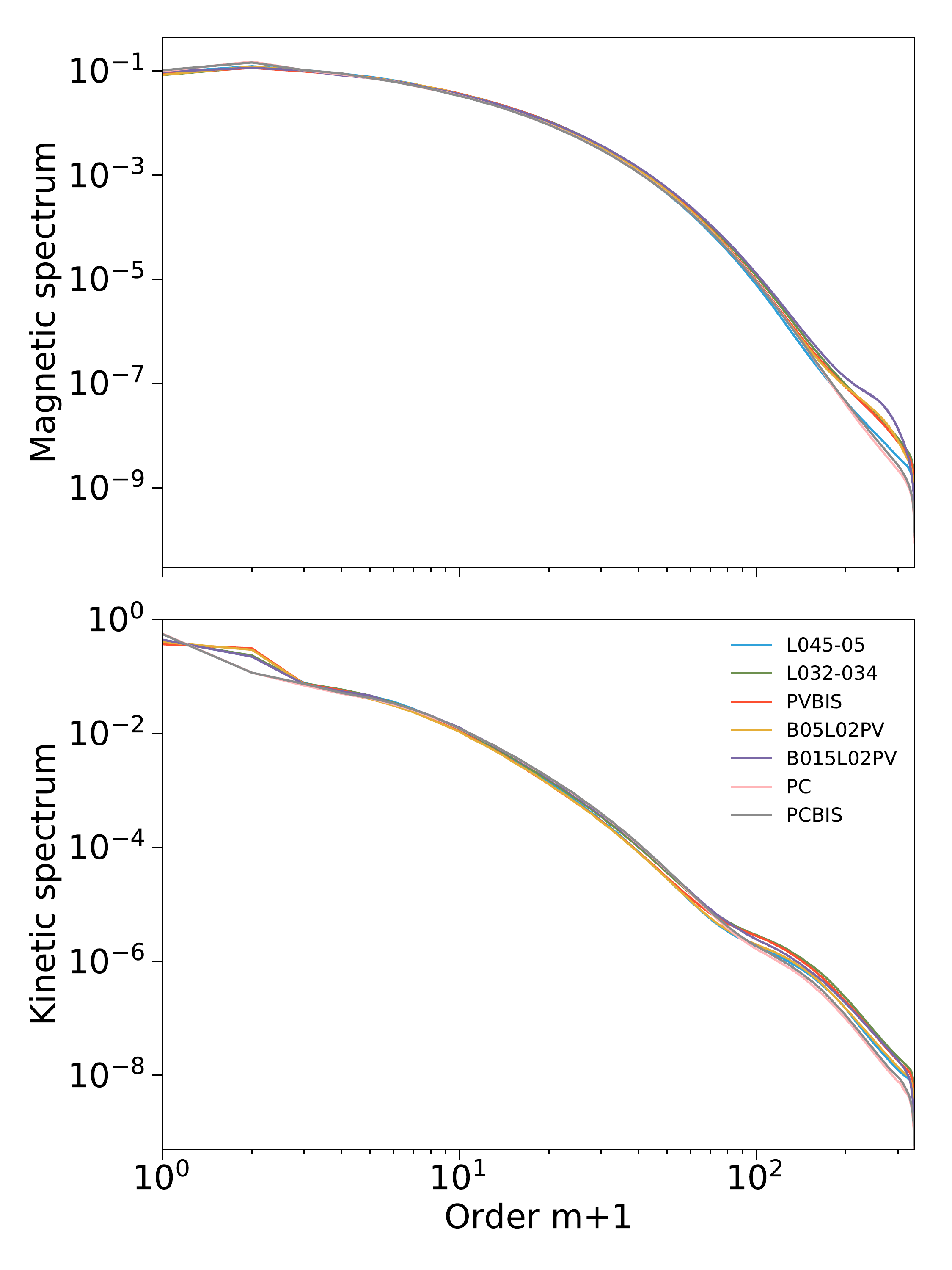}}
      \caption{Magnetic (top) and kinetic (bottom) normalized energy spectra as a function of the azimuthal degree $m$ for different models.}
         \label{Spec_m}
   \end{figure}

It is important to assess to which extent our setup impacts the magnetic field morphology, and particularly the dipole component that is constrained by observations.
The normalized magnetic and kinetic spectra depending on the Legendre polynomial order $l$ (Fig. \ref{Spec_l}) and the azimuthal degree $m$ (Fig. \ref{Spec_m}) show the differences at small and large scales for different models.
The magnetic and kinetic spectra coincide well for the intermediate and small scales (top and bottom panel of Fig. \ref{Spec_l}), while the magnetic spectrum displays significant stochasticity at the large scales.
For example, some models have a higher magnetic quadrupole $l=2$, while other models have a maximum for a higher order $l$. It seems also that the models with perfect conductor boundary conditions (\texttt{PC} and \texttt{PCBIS}) have a lower kinetic $l=1$ mode.
As described in Sect.~\ref{MRI_example}, the oscillations in the kinetic spectra are due to the meridional circulation.
The magnetic spectra as a function of the azimuthal degree $m$  (Fig. \ref{Spec_m}) is dominated by the large scales with a constant energy for $m\lesssim 10$ and decreases sharply for $m\gtrsim 10$.
If we compare the $l$ and $m$ spectra, we note that the dominating scales are larger for the azimuthal spectrum, which is consistent with the elongated structures we can see in the equatorial plane of  Fig.~\ref{3D_figure}.
Similarly, the kinetic energy decreases more steeply as a function of the azimuthal degree $m$ than as a function of $l$.
All models show a very good agreement for these kinetic and magnetic spectrum.

\begin{figure}
  \centering
   \resizebox{\hsize}{!}
            {
     \includegraphics[width=6cm]{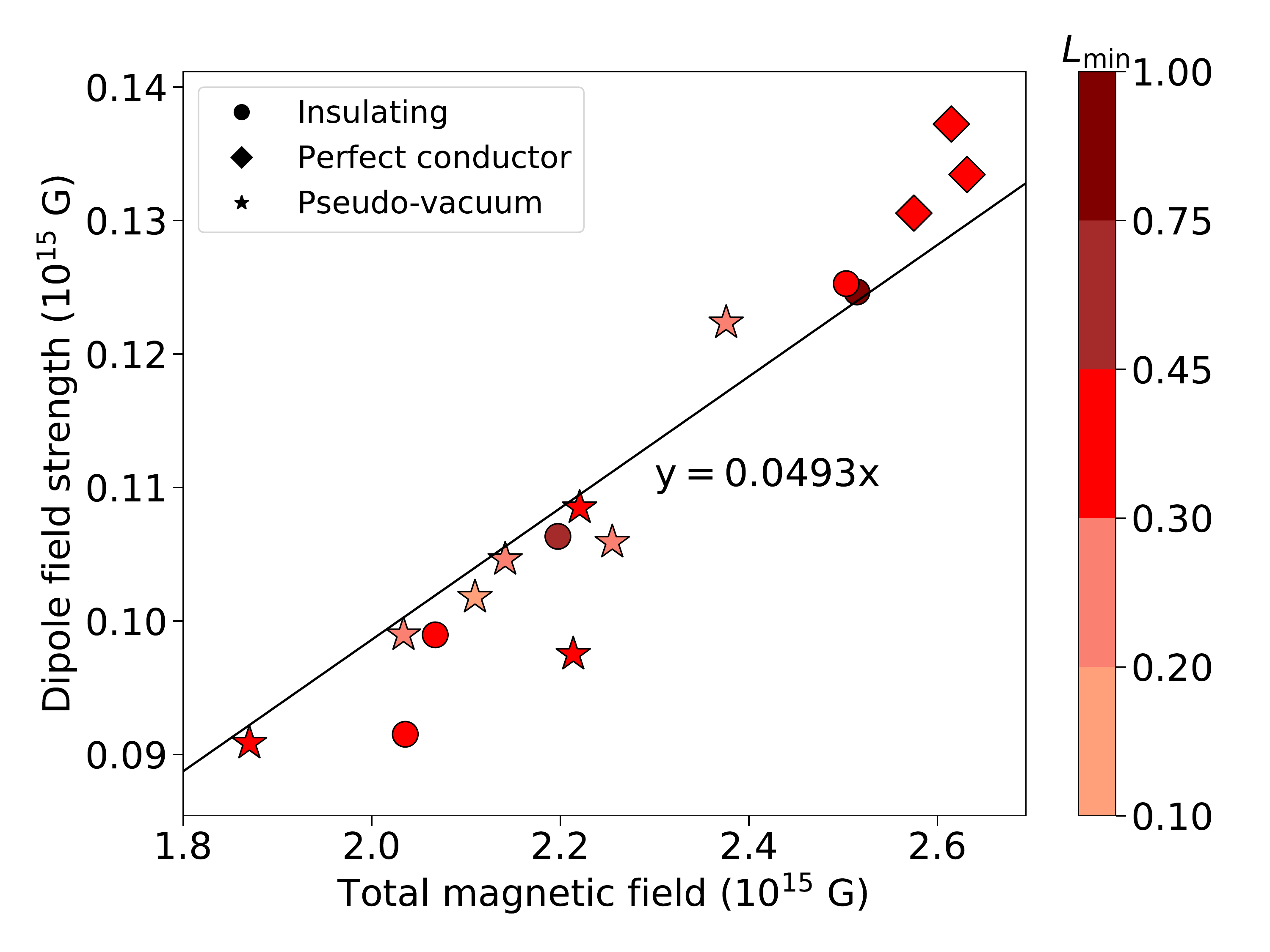}}
      \caption{Dipole magnetic field strength as a function of the averaged magnetic field strength. Symbol are defined in the caption of Fig.~\ref{max_amp_B0}.}
         \label{Bdip_Btot}
   \end{figure}

Even though the magnetic spectra of different simulations show significant dispersion at large scales, the dipole amplitude scales linearly with the averaged magnetic field (Fig. \ref{Bdip_Btot}).
The generation of a dipolar magnetic field is therefore a robust feature of the MRI and both initial and boundary conditions have a small impact on our qualitative and quantitative results, such as the dipole and the total magnetic field.

\section{Comparison to local simulations}
\label{local_comp}

\begin{figure}
  \centering
   \resizebox{\hsize}{!}
            {
     \includegraphics{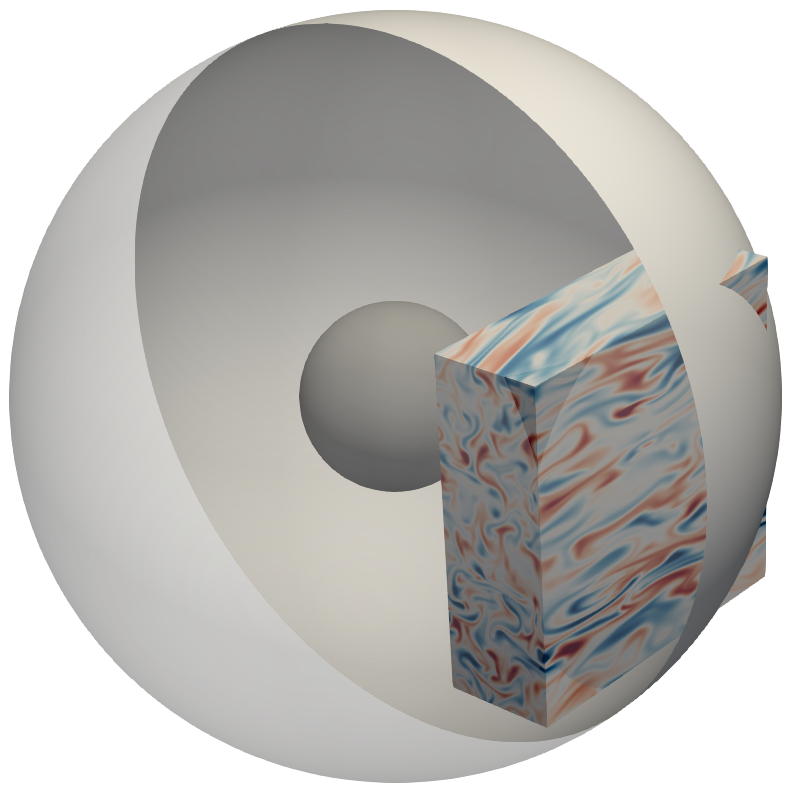}}
      \caption{Toroidal magnetic field $B_{\phi}$ of the local model \texttt{Ls03Lp09Lz09}  with a box size $(L_s,L_{\phi},L_z)=(0.3 r_\mathrm{o},0.9 r_\mathrm{o},0.9 r_\mathrm{o})$ compared to the spherical domain of our global model.}
         \label{Local2global}
   \end{figure}

In this section, we would like to understand the impact of the geometry of the domain on the properties of the MRI.
As our study is one of the first global spherical models that resolves the MRI, it is important to compare our results to local model results.
We intend to establish whether a local Cartesian box can reproduce faithfully some of the properties of the MRI in a global spherical shell. A schematic view of the comparison is shown on Fig. \ref{Local2global}.
The Cartesian box is meant to represent the differentially rotating zone in the equatorial region of our global model.
The Cartesian coordinates (x, y, z) in the local model corresponds to the cylindrical coordinates (s, $\phi$, z) at the equator of the global model. The center of the box is at a radius of $s_0 \sim 0.7 r_\mathrm{o} $.
For the cylindrical radii $s_\mathrm{cyl} \in [0.525 r_\mathrm{o},0.9 r_\mathrm{o}]$, the shear rate $q$ is approximately constant and takes varying values in the range $[0.7, 1.0]$ for different models (see Fig. \ref{shear_profile}).

\subsection{Setup for local models}

The local models were computed with the code SNOOPY, which has been widely used to study the MRI \citep[and references therein]{2005A&A...444...25L,2007MNRAS.378.1471L,2015MNRAS.450.2153G}.
The rotation rate $\Omega_0$ and the diffusivities in the local model are equal to the values used in the global model.
The initial magnetic field was initialized with a sinusoidal radial profile of vertical field, which has been chosen such that the magnetic flux vanishes similarly to the global model.
We have checked that the quasi-stationary turbulence is independent of the initial condition, provided that the magnetic field is strong enough to initiate an MRI dynamo.

To investigate the effect of the size of the domain and its geometry, we chose different box sizes $(L_s,L_{\phi},L_z)$.
If the properties of the MRI dynamo were independent of the domain geometry, then we would expect similar results when models have the same volume.
The box $(L_s,L_{\phi},L_z)=(0.5 r_\mathrm{o}, 4.5 r_\mathrm{o}, 1.2 r_\mathrm{o})$ matches the volume of the most turbulent region $V_\mathrm{turb}$ in the global model where differential rotation is strong, which corresponds to a cylindrical radius $s_\mathrm{cyl} \ge 0.5 r_\mathrm{o}$.
Models with reduced box dimensions, especially the azimuthal length $L_{\phi}$, have also been considered in an attempt to obtain results more similar to the global models.
The volume of the box for the different box dimensions considered ranges from $V=0.04 V_\mathrm{turb}$ to $V= V_\mathrm{turb}$ (see Table \ref{table:1}).

\subsection{Dynamo threshold}
\label{local_thresh}

\begin{figure}
  \centering
   \resizebox{\hsize}{!}
            {
     \includegraphics[width=9cm]{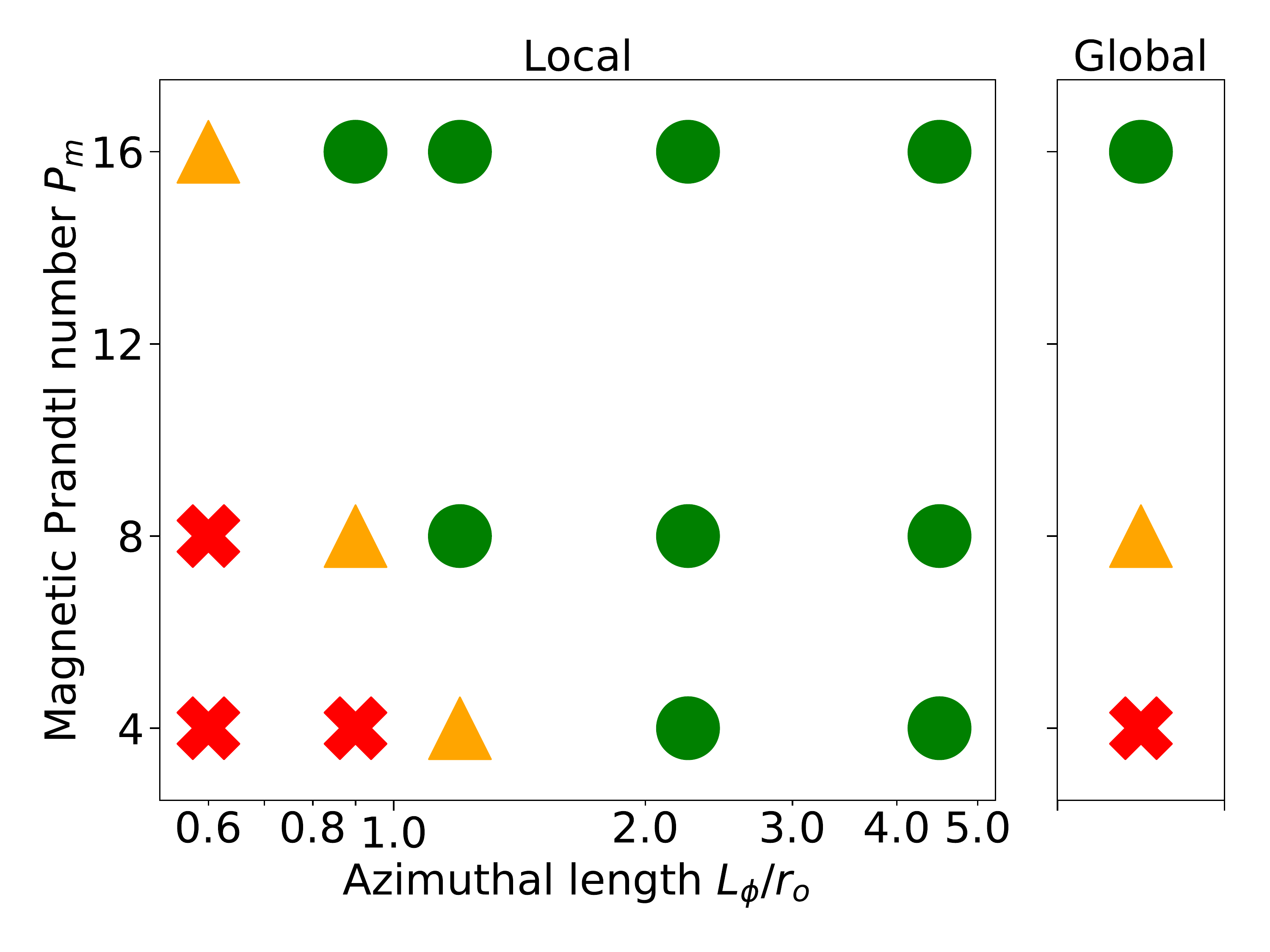}}
      \caption{Summary of the state (dynamo or not) of the model in a $(L_{\phi},P_\mathrm{m})$ plane for both local and global models. A green circle corresponds to a self-sustained dynamo. An orange triangle corresponds to a transient dynamo. A red cross corresponds to no dynamo.}
         \label{dynamo_plot}
   \end{figure}

For comparison with the results of Sect. \ref{Threshold}, we determined the threshold in magnetic Prandtl number $P_\mathrm{m}$ necessary for dynamo action as a function of the azimuthal length $L_{\phi}$.
Figure \ref{dynamo_plot} compares the dynamo threshold in the global model (right part) and the local model (left part).
For the global model we obtain no dynamo for $P_\mathrm{m} \leq 4$, a transient dynamo for $P_\mathrm{m}=8$ and a stable dynamo for $P_\mathrm{m} \geq 16$.
The local model that matches this global behavior corresponds to a box size of $(L_s,L_\phi,L_z)=(0.3r_\mathrm{o},0.9r_\mathrm{o},0.9r_\mathrm{o})$ and the $P_\mathrm{m}$ threshold decreases for larger boxes.
This shows that the dynamo threshold is higher in the global model than the local model with the same volume.

\subsection{Turbulent energies and angular momentum transport}
\label{Comp}

To make a more detailed comparison between the local and the global models, we now consider the turbulent transport of angular momentum as well as the kinetic and magnetic energies.
For comparison with the local models, the quantities from the global models were averaged for the cylindrical radius in the interval $s_\mathrm{cyl} \in [0.525 r_\mathrm{o},0.925 r_\mathrm{o}]$.
We excluded a thin layer of thickness $0.075 r_\mathrm{o}$ near the outer boundary, where the turbulence is strongly affected by the boundary conditions.

\begin{figure}
  \centering
   \resizebox{\hsize}{!}
            {
     \includegraphics[width=6cm]{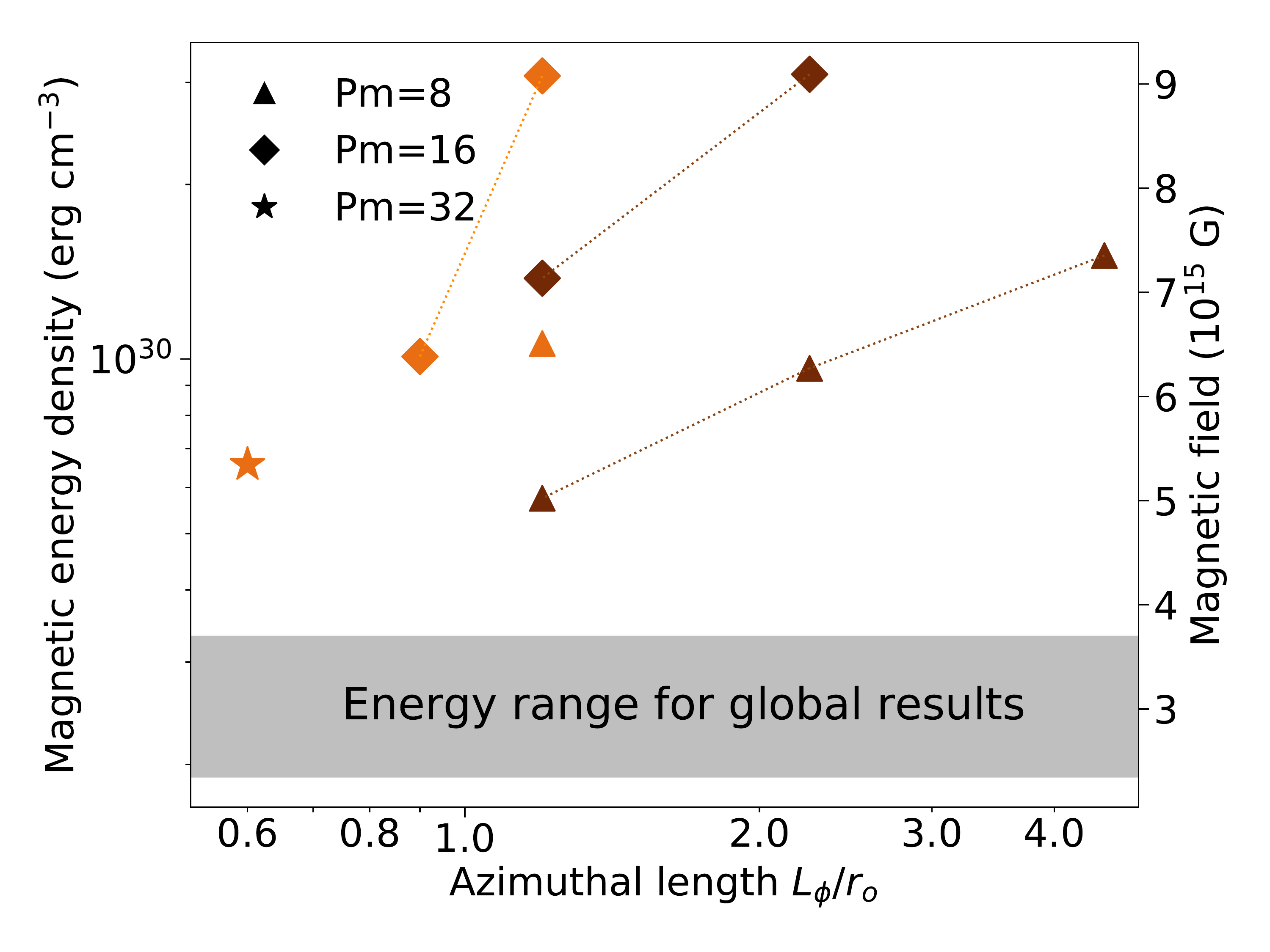}}
      \caption{Averaged magnetic energy density as a function of the box azimuthal size. The symbol shape represents the magnetic Prandtl number $P_\mathrm{m}$. The symbol color indicates the radial length of the box (brown: $L_s=0.67$; orange: $L_s=0.4$). The gray zone represents the range of magnetic energy density averaged in the turbulent zone for our sample of global models.}
         \label{Comp_Ly}
   \end{figure}

Figure \ref{Comp_Ly} shows the magnetic energy density as a function of the azimuthal length $L_{\phi}$ for different magnetic Prandtl number $P_\mathrm{m}$, and different box lengths $L_s$ and $L_z$.
All local models have a significantly higher energy density than the global models (gray area), with the smaller boxes having a magnetic energy density closer to the global models.
The impact of the box dimensions can also be deduced from these results.
First, for a same radial length $L_s$ (same color) and $P_\mathrm{m}$ (same marker), the magnetic energy density increases with the azimuthal length $L_{\phi}$.
Moreover, the magnetic energy density increases with the magnetic Prandtl number $P_\mathrm{m}$, which was expected from previous local MRI studies \citep{2007MNRAS.378.1471L,2007A&A...476.1123F,2010A&A...516A..51L}.
Figure~\ref{Comp_Ly} shows that the global simulations tend to have a lower averaged magnetic energy density than the local simulations.

In MRI turbulence, ratios of different time averaged quantities can be more robust, showing less variation than the turbulent energies. For example, the ratio of the magnetic to kinetic turbulent energies is around $10$ for both local and global models (see Table \ref{table:1}). One can also look at the turbulent stresses and the different component contributions to the turbulent energies to compare in more details the local and global models (Table \ref{table:1}).
For this diagnostic, we used the $s\phi$ component of the Maxwell and Reynolds stress tensors,
respectively defined as
\begin{align}
    M_\mathrm{s\phi} &= -\frac{1}{\rho_0 \mu_0 }\big \langle \frac{B_s \ B_{\phi}}{D^2\Omega_0^2} \big \rangle \,, \\
    R_\mathrm{s\phi} &= \big \langle \frac{u_s \ u_{\phi}}{D^2\Omega_0^2} \big \rangle \,,
\end{align}
where the brackets $\langle \cdot \rangle$ corresponds to the volume averaged value.
\begin{table*}
\renewcommand{\arraystretch}{1.2}
\caption{Local quantities used as diagnostics of the turbulence in both local and global models. The Maxwell and Reynolds stresses are normalized by the magnetic energy.}
\label{table:1}
\centering
\resizebox{\hsize}{!}{
\begin{tabular}{l r c r c c c c c c c c c c}  
\hline\hline
Name&$P_\mathrm{m}$&$q_\mathrm{avg}$&$(L_s,L_{\phi},L_z)$&$E_\mathrm{kin} $ &$E_\mathrm{mag}$&$M_\mathrm{s\phi}$&$R_\mathrm{s\phi}$&$u_s^2/2$&$u_{\phi}^2/2$&$u_z^2/2$&$b_s^2/2$&$b_{\phi}^2/2$&$b_z^2/2$\\
Units & - & - &[($r_\mathrm{o}$,$r_\mathrm{o}$,$r_\mathrm{o}$)] &[\SI{}{erg \per \cm \cubed}]  &[\SI{}{erg \per \cm \cubed}]& - & - & [\%] & [\%] & [\%] &[\%] &[\%] &[\%]\\
\hline
\textbf{LOCAL MODELS}& &  & & &&& & & & & & & \\
\hline
\texttt{Pm8Ls05Lp45Lz12}&8&0.8&(0.5, 4.5, 1.2)&1.69e29&1.51e30&0.422&0.0225&28.2&51.5&20.3&9.3&87.8&2.9\\
\texttt{Pm8Ls05Lp22Lz12}&8&0.8&(0.5, 2.25, 1.2)&9.16e28&9.64e29&0.406&0.0210&26.1&50.9&23&8.3&89.1&2.6\\
\texttt{Pm8Ls05Lp12Lz12}&8&0.8&(0.5, 1.2, 1.2)&4.85e28&5.75e29&0.381&0.0185&23.4&50.5&26.1&6.9&90.9&2.2\\
\texttt{Pm8Ls03Lp12Lz12}&8&0.8&(0.3, 1.2, 1.2)&1.10e29&1.06e30&0.405&0.0220&24&52.9&23.1&9.0&88&2.9\\
\texttt{Ls05Lp22Lz12}&16&0.8&(0.5, 2.25, 1.2)&3.51e29&3.10e30&0.439& 0.0217&24&57.8&18.2&11.75&84.3&3.95\\
\texttt{Ls05Lp12Lz12}&16&0.8&(0.5, 1.2, 1.2)&1.29e29&1.38e30&0.433&0.0215&23.6&54.8&21.6&9.65&87.2&3.15\\
\texttt{Ls03Lp12Lz12}&16&0.8&(0.3, 1.2, 1.2)&2.53e29&3.08e30&0.356& 0.0185&23.2&60.4&16.4&9.75&87.4&2.85\\
\texttt{Ls03Lp09Lz09}&16&0.8&(0.3, 0.9, 0.9)&8.25e28&1.01e30&0.411&0.0205&23&55.3&21.7&8.9&88.2&2.9\\
\texttt{Pm32Ls03Lp08Lz08}&32&0.8&(0.3, 0.8, 0.8)&3.95e28&6.57e29&0.389&0.0168&22.7&55.5&21.8&7.05&90.6&2.35\\
\hline
\textbf{GLOBAL MODELS}& & & & &&& & & & & &\\
\hline
\texttt{PVBIS}&16&0.977&-&1.70e28 & 1.71e29&0.301&0.0152&22.7&55.8&21.5&4.05&94.2&1.75\\
\texttt{PCBIS}&16&0.819&-&1.86e28 & 3.11e29&0.299&0.00932&28.1&46.4&25.5&4.2&93.2&2.6\\
\texttt{StandardBIS}&16&0.95&-&1.56e28& 1.92e29&0.302&0.0130&25.3&51.1&23.6&4.1&94&1.9\\
\texttt{B04L015PV}&16&0.951&-&1.28e28 &1.90e29&0.305&0.0116&28.7&44.4&26.9&4.1&94.1&1.8\\
\texttt{B05PV}&16&0.903&-&1.69e28 & 2.29e29&0.304&0.0120&26.2&49&24.8&4.2&93.7&2.1\\
\texttt{B08L015PV}&16&0.859&-&1.90e28 & 2.65e29&0.325&0.0114&27.5&46.8&25.7&4.9&92.9&2.2\\
\texttt{B05L008-023PV}&16&0.936&-&1.73e28&2.11e29 &0.313&0.0135&25.4&50.1&24.5&4.45&93.5&2.05\\
\texttt{B05PC}&16&0.793&-&2.42e28 &3.33e29 &0.304&0.0108&25&52.7&22.3&4.5&92.8&2.7\\
\hline
\end{tabular}}
\end{table*}
The Maxwell and Reynolds stresses normalized by the magnetic energy are weaker in global models, which shows that the angular momentum transport is less efficient.
This difference can be explained for the Maxwell stress by the lower value of the magnetic radial contribution $b_\mathrm{s}$, while the other contributions are similar in proportion. The difference is also reduced for the local models that have a low magnetic energy.

\begin{figure}
\centering
\resizebox{\hsize}{!}
            {
     \includegraphics{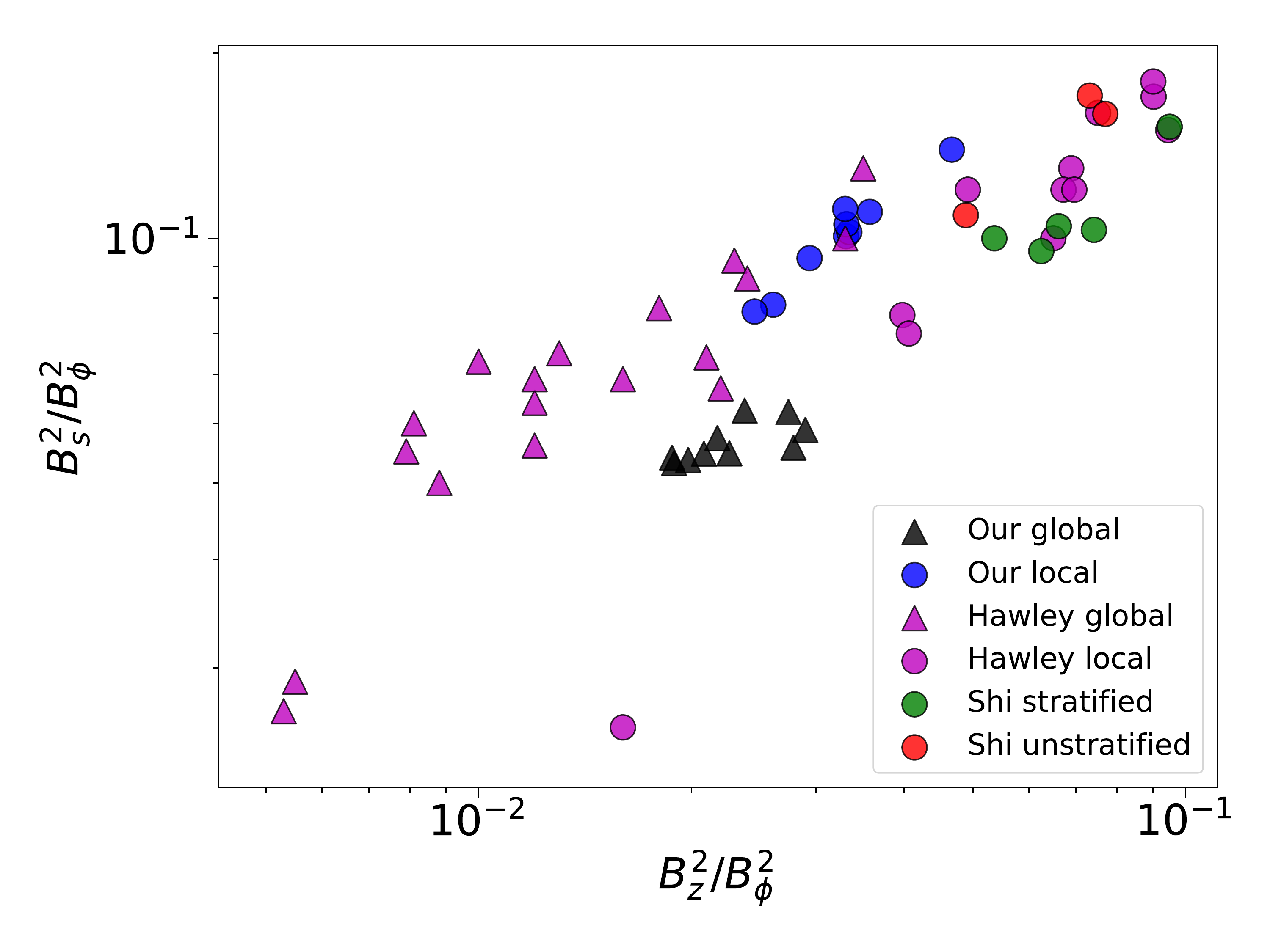}}
      \caption{Comparison for different models of the radial component $B_s^2$ as a function of vertical component $B_z^2$. Both components are normalized by the azimuthal component $B_{\phi}^2$. The different values are taken from \citet{2011ApJ...738...84H} and \citet{2010ApJ...708.1716S,2016MNRAS.456.2273S}.}
         \label{Ratio_comp}
\end{figure}

Figure \ref{Ratio_comp} shows that there is a correlation between the radial and vertical components of the magnetic field for physically different models, such as our local and global incompressible models, compressible stratified local and global models from an accretion disk study \citep{2011ApJ...738...84H} and stratified and unstratified local models \citep{2010ApJ...708.1716S,2016MNRAS.456.2273S}.
Our local models seem to match well this trend, while our global models have a slightly lower radial contribution.
It is unclear whether this small difference hints at an underlying systematic effect or is instead consistent with an inherent spread in the distribution of the data.

\begin{figure}
\centering
 \resizebox{\hsize}{!}
            {
     \includegraphics[width=6cm]{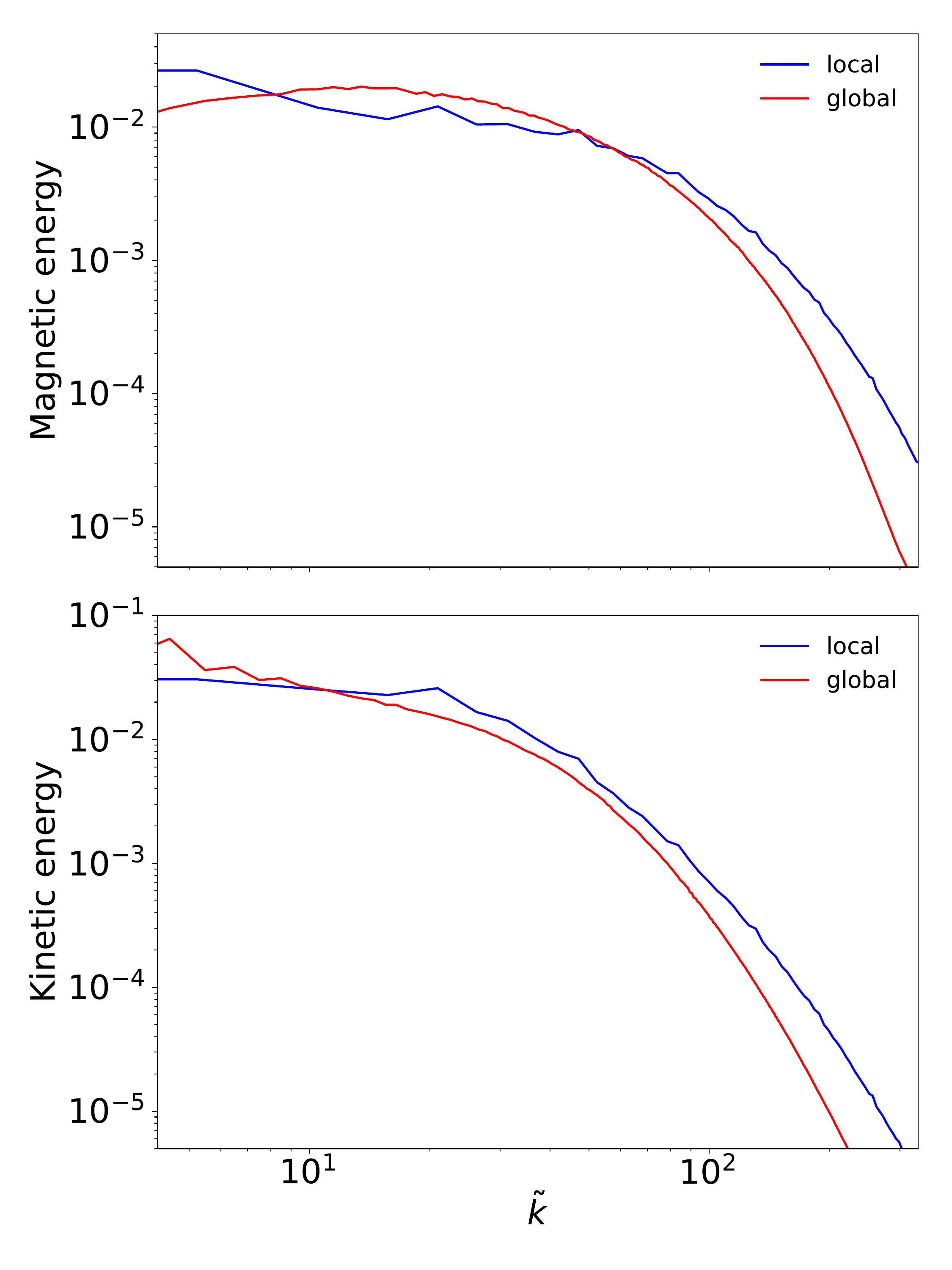}}
      \caption{Spectrum of the magnetic energy and kinetic energy as a function of the wavenumber $\tilde{k}$. The blue spectrum is averaged on the full duration of the local model \texttt{Ls03Lp12Lz12} and the red spectrum is the total nonaxisymmetric spectrum of the global model \texttt{Standard} (Fig. \ref{standard_spectrum}).}
         \label{Local_spec_l}
\end{figure}

Among local models, \texttt{Ls03Lp09Lz09} is the one showing features most similar to the global models (see the 3D snapshot in Fig. ~\ref{Local2global}  and the one shown in Fig. ~\ref{3D_figure} for a comparison).
In addition to sharing the same threshold in $P_\mathrm{m}$ for dynamo action (Sect.~4.2), it has the smallest magnetic energy among the local models at $P_\mathrm{m}$=16.
Its kinetic and magnetic spectra can therefore be compared to the global nonaxisymmetric spectra (Fig.~\ref{Local_spec_l}).
The non-radial wavenumber $\tilde{k}$ is related to the spherical harmonic order $l$ by the relation $\tilde{k} = \sqrt{{l(l+1)}/{r^2}}$ with $r$ taken as $r=0.75 r_\mathrm{o}$ for the global model in order to simplify the comparison.
The magnetic spectrum of the local model is similar to the global spectrum.
It can also be decomposed in an approximately flat part at large scales and an exponential decrease due to dissipation at small scales.
The local kinetic spectrum is also in good agreement with the nonaxisymmetric spectrum global model.
The local magnetic and kinetic spectra are stronger than the global ones at small scales, which is consistent with the stronger turbulence in our shearing boxes leading to slightly smaller dissipative scales.
The main difference between the global and local spectra is therefore the presence of scales larger than the box, especially the axisymmetric component of the kinetic spectrum. Overall, these results suggest that the self-sustained dynamo obtained in the global models is typical of the MRI in local simulations.

\section{Discussion}
\label{discuss}

In this section, we discuss the influence of the domain geometry and the different limits of the simplifying assumptions made in our study: the impact of the curvature, the mechanical boundary conditions, the diffusive processes and the incompressible approximation.

\subsection{Impact of the curvature}
\label{loc_glob}
As the local and global models are designed to have a similar differential rotation, one might have naively expected to obtain quantitatively similar results.
However, the different geometry of the domain is sufficient to obtain different results between the local and global models.
Our interpretation is that the curvature of the sphere and non-periodic boundaries reduce the coherence of the magnetic field and velocity field.
Thus, local models and large boxes in particular tend to overestimate the field amplification, since they allow for larger coherence structures to develop and thus favor the MRI action over a broader range of scales.

\subsection{Forcing of the differential rotation}

We chose to rely on the outer mechanical boundary to force the differential rotation in the global simulations, which is justified by the dynamical evolution of the PNS in core-collapse supernovae.
Indeed, the PNS contraction accelerates its rotation.
At the same time, the MRI and the MHD turbulence transport the angular momentum toward the outer cylindrical radii and therefore slow down the rotation.
This behavior may even last longer than the duration of the core-collapse supernova because the contraction of the PNS can still continue after the explosion and the rotation can slow down due to winds or magnetic braking.
These processes lead to the formation of two distinct regions within the PNS: The inner cylindrical part of the PNS is accelerated, while the outer cylindrical part is slowed down by the angular momentum transport.
Our boundary conditions try to mimic this effect. We note that, despite being fixed at the boundary, the rotation profile inside the domain is free to evolve as a consequence of the backreaction of the magnetic field. As illustrated in Figure~\ref{shear_profile}, we indeed observe that a stronger magnetic field leads to a lower degree of differential rotation.

However, one of the main limitations of this boundary condition is that the evolution of the global rotation profile is not well described as the effects of contraction, accretion, magnetic braking and transport of angular momentum is artificially taken into account on the outer boundary.
Indeed, the timescale over which the rotation profile is significantly changed can be estimated by the amount of shear energy extracted by the combined effect of Reynolds and Maxwell stresses in the turbulent region.
We estimate a shear energy of $3.9 \times 10^{50}$ erg by comparing the rotational energy contained in the differentially rotating PNS to that of a PNS in solid body rotation with the same total angular momentum.
The typical mean rate at which the shear energy is extracted is $\approx 5.2 \times 10^{51} $ erg s$^{-1}$.
The typical timescale at which the energy is extracted from the shear is therefore $\approx 75$ ms.
This means that differential rotation would decrease rapidly without contraction and accretion.
In MHD core collapse simulations, it remains unclear whether the speed-up due to contraction and accretion or rather the action of stresses and magnetic braking have a stronger impact on the rotation profile.

The energy injection by this outer boundary condition also depends on the viscous layer, and therefore on the diffusion processes.
In addition, the energy injection of the outer boundary is different from the constant shear used in local models, which might be an other way to explain the differences between local and global models.

\subsection{Diffusive processes} \label{Diffusion}

One could be tempted to solve our set of equations in the ideal MHD limit. However, resolution studies have shown that the turbulent transport is then sensitive to the numerical resolution \citep{2007ApJ...668L..51P}.
Therefore, in order to avoid the pitfall of resolution-dependent results, we used explicit diffusivities and made sure that the diffusive scales were resolved. The low numerical diffusion of the pseudo-spectral codes ensure that only physical diffusivities impact the dynamics.

The regime of the diffusive parameters that can be numerically reached is a limit to our model: A realistic value of the neutrino viscosity is $\nu \sim 2\times10^{10}$ cm$^2$ s$^{-1}$ at a radius of $\sim$\SI{20}{\km} \citep{2015MNRAS.447.3992G}. The corresponding Ekman number is then
$E \sim 6\times 10^{-6}$,
which is lower than in our simulations by a factor of about 30. Using a more realistic value of $E$ may lower the  thresholds to obtain a dynamo (see Sect. \ref{Threshold}) and may lead to a stronger amplification of the magnetic field.
Moreover, the resistivity used in our simulations is much larger than realistic values. Due to numerical constraints, the magnetic Prandtl number is limited to 16 in our study, much smaller than realistic values of $P_\mathrm{m}\sim 10^{13}$.
Accretion disk studies have shown that the MRI-driven turbulence is highly sensitive to the magnetic Prandtl number, and generally it increases with $P_\mathrm{m}$
\citep{2007MNRAS.378.1471L,2007A&A...476.1123F,2010A&A...516A..51L}.
Our local numerical simulations confirm this trend (see Table \ref{table:1}).
Our quantitative results should therefore be considered as lower bounds on the magnetic field strength that would be reached at higher $P_\mathrm{m}$.
Investigating the regime of higher $P_\mathrm{m}$ in a global model is computationally demanding and it would certainly be more affordable to use local models.
Further comparisons between local and global simulations would then allow us to extrapolate the results to a global model.

Finally, we assume that neutrinos are in the diffusive regime, which should be valid only in the inner part of the PNS.
In the outer parts of the PNS, the neutrinos are in the nondiffusive regime on length scales at which the MRI grows. The impact of the neutrino drag regime on the linear phase has been studied by \citet{2015MNRAS.447.3992G}, but has never been studied in nonlinear numerical simulations. The evolution of the MRI in this regime remains an open question.

\subsection{Incompressible approximation}

The incompressible approximation was used to provide an idealized reference model and to reduce the cost of our numerical simulations by filtering out sound waves.
First, filtering sound waves is justified when the sound speed $c_\mathrm{s}$ is much larger than both the fluid and Alfv\'en velocities ($v_A\equiv B/\sqrt{\mu_0 \rho}$).
With $c_s \sim $ \SI{5e4}{\km \per \second} at $r\sim \SI{20}{\km}$,
we check a posteriori that this condition is satisfied in our global models, for which ${u^2}/{c_s^2} \leq {v_A^2}/{c_s^2} \leq 10^{-4}$. This is consistent with the discussion of the Boussinesq approximation in \citet{2015MNRAS.447.3992G}.
Second, we neglect the composition and entropy gradients. The incompressible approximation could be extended to take into account the buoyancy within the Boussinesq approximation.
However, we do not include it for the sake of simplicity and in order to be able to compare our results to most of the numerical studies of the MRI in the literature.
Finally, density stratification is neglected and its study is postponed to a further work.

\section{Conclusions}
\label{conclusions}

For the first time, we have investigated the generation of large-scale magnetic fields by the MRI in an idealized 3D spherical model of PNS.
We performed a parameter exploration of the initial and boundary conditions
and we compared in details the results of our global simulations with local shearing box simulations.
The main findings of our study can be summarized as follows:
\begin{itemize}
   \item The MRI  leads to a quasi-stationary state with a strong turbulent magnetic field of $B \geq 2\times 10^{15} \ \mathrm{G}$.
   The toroidal component of the magnetic field dominates over the poloidal field by a factor of $\sim$2. A nondominant magnetic dipole $B_\mathrm{dip} \sim 10^{14} \ \mathrm{G}$  is generated, which represents about $5\%$ of the averaged magnetic field strength.
  Interestingly, this dipole is tilted toward the equatorial plane with a tilt angle in the interval $\theta_\mathrm{dip} \in [60^{\circ},120^{\circ}]$.
   \item The magnetic field amplification and dipole generation by the MRI is a robust mechanism that operates for a large number of different initial setups and magnetic boundary conditions. Indeed, the turbulent energies and the angular momentum transport in the quasi-stationary state do not depend much on the initial magnetic field and boundary conditions.
   \item The comparison between global and local studies suggests that the geometry of the domain leads to a decrease in the turbulent energies.
   Indeed, global models have lower turbulent energies than local models, although the ratio of kinetic and magnetic energies and the relative contributions to these energies are comparable. The results from the small boxes of the local simulations have a better agreement with those from the global models, which may be interpreted by the fact that the curvature in a spherical shell limits the coherence length scale of the turbulence.
\end{itemize}

These findings support the ability of the MRI to generate a strong magnetic dipole and explain magnetar formation. One should of course keep in mind that our results still need to be confirmed by setups with a more realistic PNS structure as discussed below.
\newline

The magnetic dipole amplitude obtained from the time and volume averages ranges from $9.1 \times 10^{13}$ G to $1.35 \times 10^{14}$ G, which is within the lower end of the observed range for the dipole of galactic magnetars. We further note that these results are obtained before the full contraction to the final size of a cold neutron star, which has a radius close to \SI{12}{\km}.
If the magnetic flux is conserved during this contraction, we expect the magnetic field to be amplified by a factor of $\simeq$4, possibly bringing the dipolar component of the magnetic field in the middle of magnetar range $10^{14}-10^{15}\rm G$.

To compare our results directly with observations, the turbulent magnetic field should also be relaxed to a stable state without differential rotation and be evolved on longer timescales. 
Under the joint effects of magnetospheric dissipation and internal dissipation, the dipole from our model could become a stable equatorial dipole \citep{2018MNRAS.481.4169L}.

Similarly to our results, some observations suggest that the dipole may not be the strongest component of the magnetic field of magnetars.
Indeed, \citet{2016PASJ...68S..12M,2019PASJ...71...15M} have detected phase modulations in X-ray emissions from two magnetars, which are interpreted as a free precession due to prolate deformations of the neutron star under the torque of a very intense internal toroidal field. The MRI-driven magnetic field is mainly toroidal and could be important to explain these observations.

As discussed above, our results suggest that a fast rotating PNS can become a typical magnetar through MRI action. A slower rotation, on the other hand, is expected to lead to a lower magnetic field and may explain the formation of low field magnetars \citep{2012ApJ...754...27R,2013ApJ...770...65R,2014ApJ...781L..17R}.
In fact, owing to the simplicity of our setup, our simulations can be rescaled to a different rotation frequency. For example, for a 30 times slower rotation rate\footnote{The Ekman number used in our simulations would then correspond to the realistic value of the viscosity $\nu=2\times10^{10}\rm\, cm^2/s$.}, the magnetic field should be scaled down by a factor of 30. The final neutron star would have a magnetic dipole of $\simeq 1-2\times 10^{13}$ G and a total magnetic field of $\simeq 3 \times 10^{14}$, which corresponds to a low field magnetar.
The ratio of the magnetic dipole to the total magnetic field can also be compared to the cases of SGR 0418+5729 \citep{2013Natur.500..312T} and J1822.3-16066 \citep{2016MNRAS.456.4145R}, for which the total magnetic field can be measured using the proton cyclotron line. The observed ratios are respectively $B_\mathrm{dip}/B_\mathrm{tot} \sim 0.0012-0.024$ and $B_\mathrm{dip}/B_\mathrm{tot} \sim 0.01-0.05$. These are roughly in agreement with our results $B_\mathrm{dip}/B_\mathrm{tot} \sim 0.05$.
The slightly higher value found in our models may be due to the high resistivity used in our simulations, since we may expect that a lower resistivity would lead to a smaller-scale magnetic field.

The magnetic field we obtain is strong enough to potentially impact the core-collapse dynamics and launch jet-driven explosions.
However, most of the studies of jet driven explosions assume a strong axial dipole field, while our study displays an equatorial dipole. Our results open the question of the impact of a tilted dipole on the dynamics of a core-collapse supernova. The dipolar tilt angle is also a key parameter to determine the amount of thermalized energy in the magnetosphere and the energy available to launch a jet \citep{2018MNRAS.475.2659M}.
Interestingly, \cite{2020MNRAS.492...58B} showed that higher order multipoles tend to decrease the efficiency of the magnetorotational launching mechanism  but can still launch a jet, which suggests that the multipoles of our magnetic field could contribute to the explosion.

We stress that the limitations of this work need to be assessed. Our quantitative results (e.g., strength of the magnetic field and its dipolar component, dipolar tilt angle) may change with the addition of the density and entropy background gradients.
A convective dynamo with fast rotation can occur in a realistic interior model of the full PNS, a mechanism that has recently been shown to be able to form magnetars \citep{2020ScienceRaynaud}.
This raises the open question of the interaction of the MRI in a stably stratified zone with an inner convective dynamo in core-collapse supernovae.
The stably stratified region unstable to the MRI is important to link the convective dynamo buried under the PNS surface and the explosion that is launched from the surface.
\newline

This study has been focused on the core-collapse supernova context but it is also relevant to binary neutron star mergers, which display similar conditions in terms of differential rotation and neutrino radiation \citep{2017MNRAS.471.1879G}.
The MRI may grow in the hypermassive neutron star resulting from the merger and the magnetic field may be amplified up to $10^{16}$ G \citep{2013PhRvD..87l1302S,2018PhRvD..97l4039K,2019Ciolfi}.
These studies start with a strong axial magnetic dipole ($ \ge 3 \times 10^{15}$ G), while in our case we start with a small-scale field and the MRI generates a strong tilted dipole. Although it is difficult to compare the results obtained with such different initial conditions, our work tends to support the ability of the MRI to generate an magnetar-like dipole in binary neutron star mergers. A detailed study with appropriate thermodynamic background state and differential rotation profile would further assess this possibility.
This would support the invoked scenario of magnetars powering an X-ray transient as the aftermath of a binary neutron-star merger \citep{2019Natur.568..198X}.

Investigating how the magnetic field of magnetars is generated in realistic conditions is a promising avenue of research as new observations will come in the multi-messenger era: The Large Synoptic Survey Telescope (LSST) will be able to observe one hundred times more supernovae \citep{2017arXiv170804058L} and the future spatial telescope SVOM will help us to localize and observe transient events in the $\gamma$-ray sky, like electromagnetic counterparts of binary neutron-star mergers \citep{2016arXiv161006892W}.

\begin{acknowledgements}
This research was supported by the European Research Council through the ERC starting grant MagBURST No. 715368. The computations were performed on the supercomputer Occigen of the CINES (Applications A0030410317 and A0050410317). We thank Thomas Gastine, Thierry Foglizzo and S\'{e}bastien Fromang for their valuable insight and expertise.

\end{acknowledgements}

%
%
\bibliographystyle{aa} 
\bibliography{biblio}

\onecolumn

\begin{appendix} 
\appendix

\section{Parameters of our global simulations}

\begin{table}[h!]
\renewcommand{\arraystretch}{1.2}
\centering
\caption{Overview of the global models. The resolution of these simulations is $(n_r,n_{\theta},n_{\phi})=(257,512,1024)$.}
\label{table:annex}
\setlength{\tabcolsep}{5pt}
\begin{tabular}{l c c c c c c c c}
\hline \hline
Name&$P_\mathrm{m}$&Init length&$B_0$&Magnetic BCs&Dynamo&Magnetic field&Dipole&Early maximum\\
Units&-&[($r_\mathrm{o}$,$r_\mathrm{o}$)] &[$10^{15}$ G]&-&-&[$10^{15}$ G]&[$10^{15}$ G]&[$10^{15}$ G]\\ \hline
\texttt{Pm4L056-75}&4.0&(0.56, 0.75)&2.0&Insulating&No&-&-&- \\
\texttt{L034-38}&16.0&(0.34, 0.38)&0.81&Insulating&Yes&2.20&0.106&3.80\\
\texttt{L056-75}&16.0&(0.56, 75)&2.0&Insulating&Yes&2.51&0.125&6.68 \\
\texttt{L024-026}&16.0&(0.24, 0.26)&0.91&Insulating&Yes&2.07&0.099&2.57 \\
\texttt{Standard}&16.0&(0.23,0.38)&0.88&Insulating&Yes&2.50&0.125&2.49 \\
\texttt{PV}&16.0&(0.23,0.38)&0.91&Radial field&Yes&2.22&0.108&2.42\\
\texttt{PVBIS}&16.0&(0.23,0.38)&0.91&Radial field&Yes&1.87&0.091&2.30\\
\texttt{B01PV}&16.0&(0.23,0.38)&0.41&Radial field&Transient&-&-&-\\
\texttt{B005PV}&16.0&(0.23,0.38)&0.20&Radial field&No&-&-&-\\
\texttt{Pm8B05L015PV}&8.0&(0.15, 0.38)&2.0&Radial field&Transient&-&-&-\\
\texttt{B05L015PV}&16.0&(0.15, 0.38)&2.0&Radial field&Yes&2.14&0.105&3.19\\
\texttt{B015L015PV}&16.0&(0.15, 0.38)&0.61&Radial field&Yes&2.26&0.106&1.54\\
\texttt{PC}&16.0&(0.23,0.38)&0.91&Perfect conductor&Yes&2.63&0.133&2.42\\
\texttt{PCBIS}&16.0&(0.23,0.38)&0.91&Perfect conductor&Yes&2.61&0.137&2.42\\
\texttt{StandardBIS}&16.0&(0.23,0.38)&0.92&Insulating&Yes&2.04&0.092&2.56\\
\texttt{B04L015PV}&16.0&(0.15,0.38)&1.6&Radial field&Yes&2.03&0.100&3.17\\
\texttt{B05PV}&16.0&(0.23,0.38)&2.0&Radial field&Yes&2.21&0.098&3.39\\
\texttt{B08L015PV}&16.0&(0.15, 0.38)&3.3&Radial field&Yes&2.38&0.122&3.95\\
\texttt{B05L008-023PV}&16.0&(0.08,0.23)&2.1&Radial field&Yes&2.11&0.102&1.66\\
\texttt{B05PC}&16.0&(0.23,0.38)&2.1&Perfect conductor&Yes&2.58&0.131&3.39 \\ \hline \hline
\end{tabular}
\end{table}

\end{appendix}

\end{document}